\documentclass[aps,showpacs,pre,twocolumn,superscriptaddress]{revtex4}
\usepackage{graphicx}
\usepackage{amssymb}
\usepackage{amsmath}

\begin{document}

\title{Evolutionary Dynamics in a Simple Model of Self-Assembly}
\author{Iain G. Johnston}
\affiliation{Rudolf Peierls Centre for Theoretical Physics, University of Oxford, 1 Keble Road, Oxford OX1 3NP, UK}
\author{Sebastian E. Ahnert}
\affiliation{Theory of Condensed Matter, Cavendish Laboratory, 
University of Cambridge, JJ Thomson Avenue, Cambridge CB3 0HE, UK} 
\author{Jonathan P. K. Doye}
\affiliation{Physical \& Theoretical Chemistry Laboratory, Department of Chemistry, University of Oxford, South Parks Road, Oxford OX1 3QZ, UK}
\author{Ard A. Louis}
\affiliation{Rudolf Peierls Centre for Theoretical Physics, University of Oxford, 1 Keble Road, Oxford OX1 3NP, UK}

\date{\today}


\begin{abstract}
We investigate the evolutionary dynamics of an idealised model for the robust self-assembly of two-dimensional structures called polyominoes. The model includes rules that encode interactions between sets of square tiles  that drive the self-assembly process. The relationship between the model's rule set and its resulting self-assembled structure can be viewed as a genotype-phenotype map and incorporated into a genetic algorithm.   The rule sets evolve under selection for specified target structures. The corresponding, complex fitness landscape generates rich evolutionary dynamics as a function of parameters such as the population size, search space size, mutation rate, and method of recombination.  Furthermore, these systems are  simple enough that in some cases the associated model genome space can be completely characterised, shedding light on how the evolutionary dynamics depends on the detailed structure of the fitness landscape.  Finally, we apply the model to study the emergence of the preference for dihedral over cyclic symmetry observed for homomeric protein tetramers.
\end{abstract}


\maketitle

\section{\label{secint} Introduction}
Self-assembly processes, in which constituent components reliably assemble into a complete structure without external control, are ubiquitous in nature, providing the means by which sophisticated biological machinery such as protein complexes are formed within organisms \cite{goodsell2004bionanotechnology}.  A key question then arises:  how did the interactions that drive these self-assembly processes evolve over billions of years to form the optimised systems we observe today \cite{pereira2006origins,hirsh2001protein, doye2004inhibition}? Bioinformatic studies of protein complexes \cite{levy2008assembly} suggest that a number of observed trends in protein quaternary structure are caused not only by the biological function under selection, but also by the details of the evolutionary dynamics.  Some of these trends have recently been explained by using computer simulations of a simple continuous patchy particle model \cite{wilber2007reversible}  for globular proteins \cite{villar2009self, andreani}.   However, such models are computationally expensive because a detailed simulation of the assembly process is required at each step in evolutionary time.    

\begin{figure}
\begin{center}
\includegraphics[width=8cm]{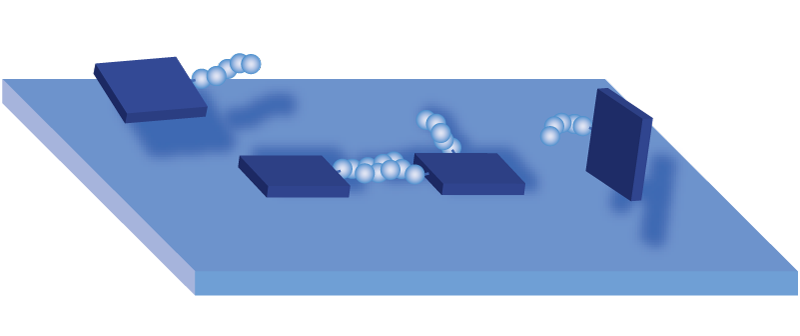}
\caption{ \footnotesize (colour online) Illustration of a possible realisation of physical polyomino assembly. Square tile building blocks interact with each other through complementary bonding between edges, here illustrated with interacting polymer chains. In addition, tiles experience an attractive interaction to a flat substrate, leading to growing polyomino structures on a surface. }
\label{real}
\end{center}
\end{figure}

In this paper we study the evolutionary dynamics of a highly idealised coarse-grained model for the evolution of self-assembling systems,  for which the assembly process can be simulated quickly and straightforwardly.  The model consists of an `alphabet' of square tiles that self-assemble into   \emph{polyominoes}: unions of connected cells on a 2D square lattice. The alphabet of available tiles, which we term the \emph{assembly rule set}, contains a description of the interactions that drive the assembling system towards a final structure \cite{AHNERT}.  A physical interpretation of the model consists of a structure assembling on a 2D substrate in contact with a suspension of tiles, as shown in Fig.~\ref{real}. 
These tiles can form many kinds of structures, both bounded and unbounded.  We focus on {\em deterministic} rule sets that always assemble into the same bounded 2D structures, a class of behaviour that is analagous to the monodisperse self-assembly observed for example for many kinds of protein quaternary structures.

These models may also be relevant for experimental systems such as 2D self-assembled systems that have been made of RNA~\cite{chwo04} and DNA~\cite{winf98} tiles.  Each tile can be tailored to interact with its neighbours through complementary bonding. Patterns and grids of varying geometries on the nanoscale have been produced by changing these design rules, with some examples being circuit patterns \cite{cook2004self} and Sierpinski triangles \cite{rothemund2004algorithmic}. The variety of structures that can be produced using DNA tiling \cite{lin2006dna} and DNA-linked particles \cite{lukatsky2006designing} is rapidly increasing.  The evolutionary design of polyomino structures may shed light on the design of these synthetic systems.

Pioneering work by Wang~\cite{ wang1963dominoes} demonstrated that tiles could be used to specify a Turing machine.  In an important development,  Winfree showed that DNA nanotechnology could be used to create molecular Wang tiles~\cite{winf98}.  Self-assembling tile sets can thus perform computational tasks such as binary counting, and  a measure of the complexity of assembly sets required for such algorithmic applications have been computed~\cite{rothemund2000program}. This theoretical work has been extended to study the details of tile assembly nucleation \cite{barish2009information} and the effects of errors in the assembly process \cite{winfree2004proofreading}.

In this study, we use genetic algorithms (GAs) \cite{goldberg1989genetic,mitchell1998introduction} that search through the space of all possible rule sets to find those that generate the deterministic assembly of desired polyomino structures.  Despite its simplicity, and resulting computational tractability, the model produces rich evolutionary behaviour.
The assembly process can be viewed as a mapping that transforms an assembly rule set into an assembled polyomino structure. This mapping is reminiscent of the \emph{genotype-phenotype map} in evolutionary biology, whereby information in the genome (the genotype) is used to develop the physical form of a biological structure (the phenotype).

We investigate how the evolutionary dynamics of our model system depends on parameters such as population size, mutation rate and recombination. In GAs, mutation rate has been shown to dramatically affect the speed of evolution, with populations evolving at higher rates around an optimal mutation rate that is roughly the reciprocal of the genome length \cite{de1990analysis, back1993optimal}. Biological organisms also often have mutation rates around this optimal value \cite{metzgar2000evidence,drake1998rates}. Recombination has also been shown to increase the speed of evolution on a simple fitness landscape \cite{cohe05}.   We study how these evolutionary variables affect the adaptation and discovery times of our self-assembling systems. 

An important property of our model is that it is simple enough to allow, in some cases,  an exhaustive search of the associated search space, yielding a fully-characterisable but highly non-trivial \emph{fitness landscape} \cite{wrig32, maynardsmith1970protein} that facilitates a detailed analysis of the underlying evolutionary dynamics. 

In addition, we aim to explore the emergence of symmetry in evolving self-assembling systems.  It has been observed, for example, that homomeric tetramer protein complexes show a strong preference for dihedral ($D_2$) symmetry over cyclic ($C_4$) symmetry~\cite{levy2008assembly,villar2009self}.  We study this preference as a function of various evolutionary parameters with our simplified polyomino system, for which a complete characterisation of the fitness landscape can be achieved.


This paper is structured as follows. In Section \ref{secmodoverall} we describe our model of self-assembling polyominoes, and our implementation of genetic algorithms. In Section \ref{secland} we exhaustively study the search space defined by a particular parameterisation of our model. Section \ref{secevdyn} analyses how evolutionary variables including mutation rate, population size and search space size affect the dynamics of polyomino evolution. In Section \ref{seclevy} we apply our model to study the evolution of homomeric tetramer protein complexes, and we list our  conclusions in Section \ref{secconc}.

. 

\section{\label{secmodoverall}Model \& Methods}
\subsection{\label{secmod} Model Implementation}
\begin{figure}
\begin{center}
\includegraphics[width=8cm]{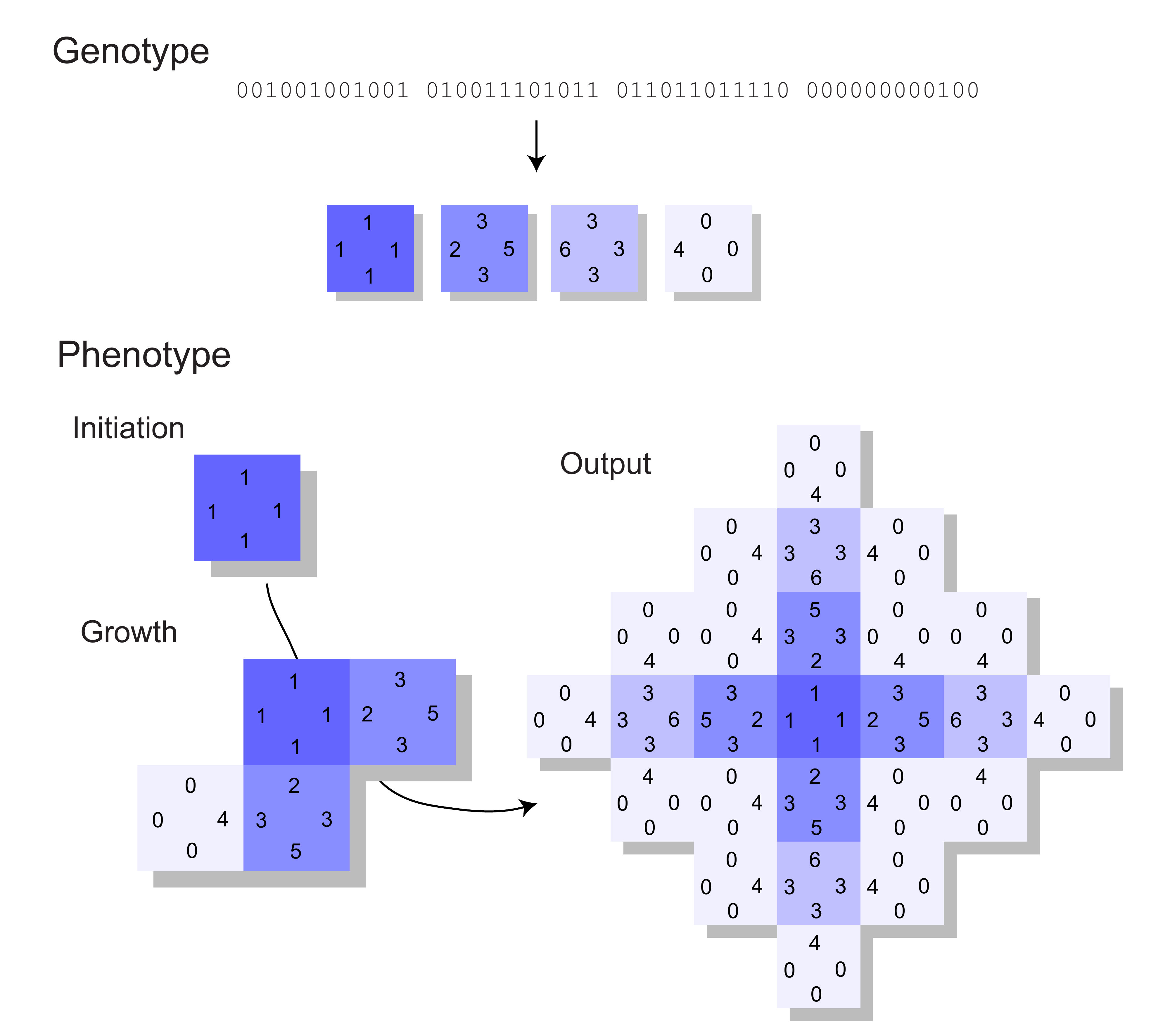}
\caption{ \footnotesize (colour online) Illustration of polyomino assembly, for a rule set with $n_t = 4, n_c = 7$, and with the nucleus and interaction conventions described in the text. A binary representation of a rule set is translated to nucleus, tile and interaction information. The nuclei are placed on a lattice in the initiation stage, and growth progresses stochastically, to a final output possessing shape and tile, but not orientational, determinism: the diagonal neighbours of the central tile have two possible orientations, as their 4 edge can bond to either of the two adjacent 3 edges.}
\label{illus}
\end{center}
\end{figure}

\begin{figure}
\begin{center}
\includegraphics[width=8cm]{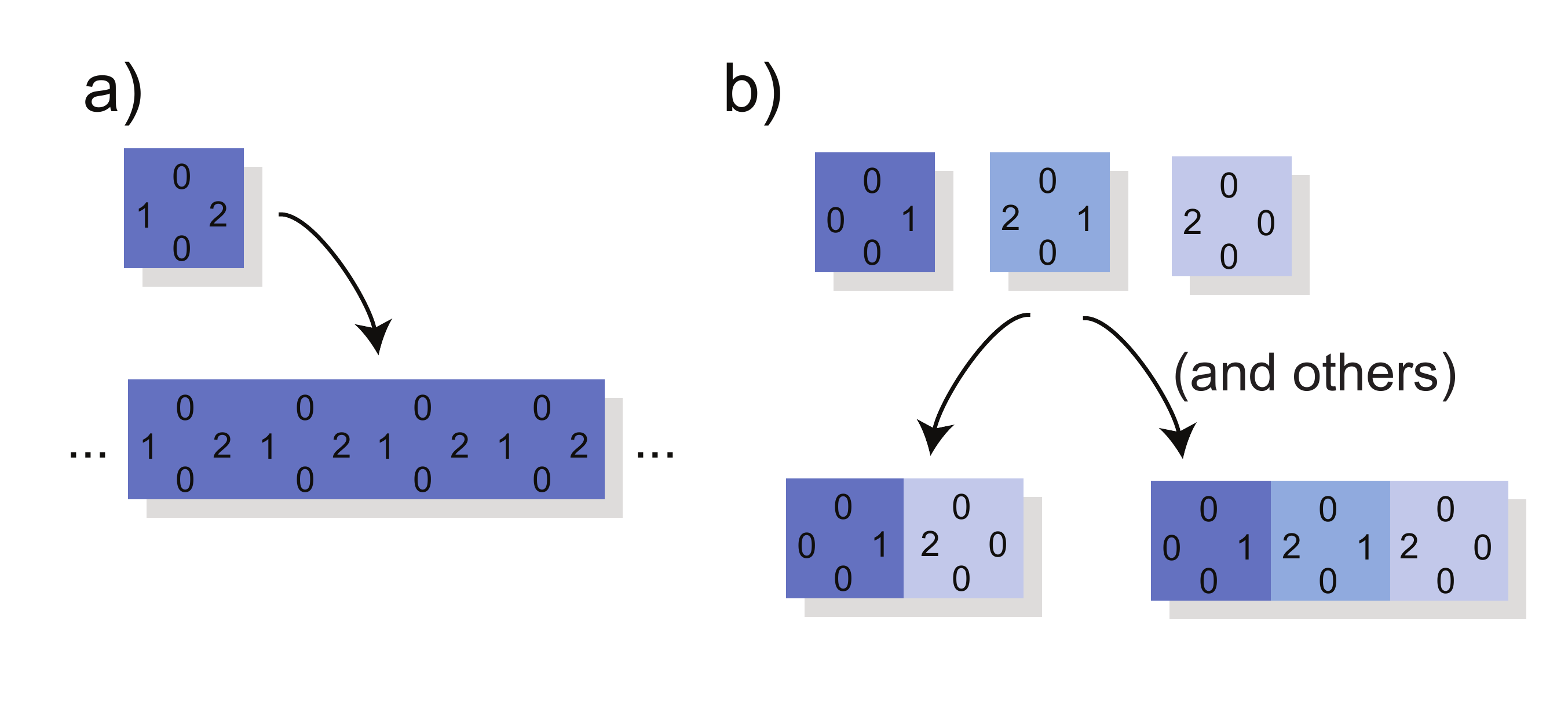}
\caption{ \footnotesize (colour online) (a) Unbound and (b) non-deterministic rule sets. The interactions in this case follow the convention in Eqn. \ref{interactions}, so edge types $1$ and $2$ experience an attractive interaction and 0 is neutral. The rule set in (a) consists of one block that is capable of bonding to itself, thus creating an endless chain of repeated blocks. The rule set in (b) contains a block capable of bonding to more than one other, leading to tile non-determinism depending on which bonding block is added first. In this case, the two different bonding tiles create different structures, leading to shape non-determinism.}
\label{undesir}
\end{center}
\end{figure}

Our model uses interacting square tiles to model the self-assembly of 2D polyomino structures on a square lattice \cite{AHNERT}. The interactions between adjacent tiles are defined by the nature of each tile's edges, which are assigned `colours', with any two colours either experiencing no interaction or an attraction. In this conceptual model, there is no energy or temperature scale, so two edges are either non-interacting or have an effectively infinite attractive interaction, making bonding irreversible. 

A given assembly scenario will consist of $n_t$ tile types, and an alphabet of $n_c$ available colours. Each tile is entirely specified by a description of its four edge colours. We will denote a tile as an ordered set of four colours, with the first element corresponding to the top edge and subsequent elements corresponding to the edges reached in clockwise order, for example, $\{1, 2, 3, 4\}$. An $n_c \times n_c$ binary interaction matrix $A$ describes the interaction between colours, with colours $i$ and $j$ experiencing an attractive interaction if $A_{ij} = 1$, and no attraction otherwise.  The generalised case of varying interaction strengths has been studied analytically \cite{solo07}, but for simplicity we consider binary interactions. 

The tiles are similar to Wang tiles \cite{wang61}, with two important differences: interactions between colours are not limited to each colour bonding only with itself, and the tiles may be rotated to any of the four possible orientations allowed by $C_4$ symmetry (for example, $\{1, 2, 3, 4\} \equiv \{3, 4, 1, 2\}$). The sides of a tile therefore comprise what is termed `an $n_c$-ary fixed necklace' of length 4 \cite{AHNERT, necklace}. The generalisation to free necklaces \cite{necklace}, in which tiles may also be `flipped' ($\{1, 2, 3, 4\} \equiv \{1, 4, 3, 2\}$ and its cyclic variants), will be visited in Section \ref{seclevy}.

Assembly progresses on an infinite square lattice, and takes places in two phases: initiation and growth (see Fig.~\ref{illus}). The initiation phase involves one or more tiles being placed on the (initially empty) lattice at prescribed positions and orientations: these are the \emph{nucleus tiles}, each of which is described by the tile type of the nucleus, its co-ordinates on the lattice, and its orientation. The combined instruction set representing nucleus, tile edge, and interaction matrix data is the \emph{rule set} for a particular assembly scenario.

There are several alternative schemes for nucleating assembly in this model. Assembly may progress from a single initial tile, laid down at the start of the assembly process. In this case, the single tile may be of a fixed, specific tile type --- which we will term a single fixed nucleus (SFN) --- or of a tile type arbitrarily chosen from the rule set --- which we will term a single general nucleus (SGN). It has been shown that to guarantee deterministic assembly from an arbitrary nucleus tile, considerably more information content is often required within genomes \cite{AHNERT}. 


The question of nucleating tile-based self-assembly has been addressed theoretically \cite{schulman2009programmable} and experimentally \cite{barish2009information} in the context of algorithmic DNA assembly. In these studies, seed particles constructed of DNA form the nucleus of a structure and contain information to regulate the assembly process. This approach effectively corresponds to an SFN setup.

We will adopt conventions for the nucleus tiles and the structure of the interaction matrix $A$, allowing us to simplify the representation of a rule set. We will use an SFN, and take the nucleus tile to be of the tile type first described in the rule set. Furthermore, we fix the orientation of the nucleus tile, so that the edge specified first in the rule set is taken to be the upper edge of the tile when first placed on the grid. Under our convention, the position of the nucleus tile is arbitrary, and polyominoes that differ only by translations are counted as equivalent. 



We will usually (with an exception in Section \ref{seclevy}, which allows the incorporation of self-interacting colours) fix the interaction matrix by defining the interaction between colours $i$ and $j$ (represented by non-negative integers) as:

\begin{equation}
A_{ij} = (1 - i\,\mbox{mod}\,2)\delta_{i (j+1)} + (i\,\mbox{mod}\, 2) \delta_{i (j-1)}
\label{interactions}
\end{equation}
so that each colour only interacts with one partner, $1 \leftrightarrow 2$, $3 \leftrightarrow 4, ...$ and 0 provides a neutral edge, which does not interact with any other edge type. 

The combination of conventions for assembly nucleation (SFN, with the first tile specified in the rule set as the nucleus) and the interaction matrix (Eqn.~\ref{interactions}) allows us to represent a given rule set by specifying the edges of the tiles involved in assembly alone. Rule sets can then be represented straightforwardly by a binary string (see Fig.~\ref{illus}), by writing each numerical parameter in the rule set (each tile edge) as its binary counterpart and concatenating all the binary variables into one long string. This resulting `genome' is then suitable for processing with genetic algorithms (see Section \ref{secga}). 

Growth progresses stochastically in the following manner. A tile type is chosen randomly from a uniform distribution over the available tiles. A position on the lattice is selected randomly with the constraint that it must be adjacent to a previously laid tile. The chosen tile is cycled in random order through its four possible orientations at the chosen point. If during this cycling the tile experiences an attractive interaction to any of its four neighbouring lattice points, it bonds immediately in that configuration at the chosen site, as illustrated in Fig.~\ref{illus}. In this way, bonding occurs irreversibly, but the model can be generalised to allow reversible interactions by introducing a temperature scale, relaxing the binary constraint on interaction matrix $A$, and allowing assembly to proceed within a simulation that includes thermal effects.

\subsection{\label{secout} Classes of Assembly Behaviour}

\begin{figure}
\begin{center}
\includegraphics[width=8cm]{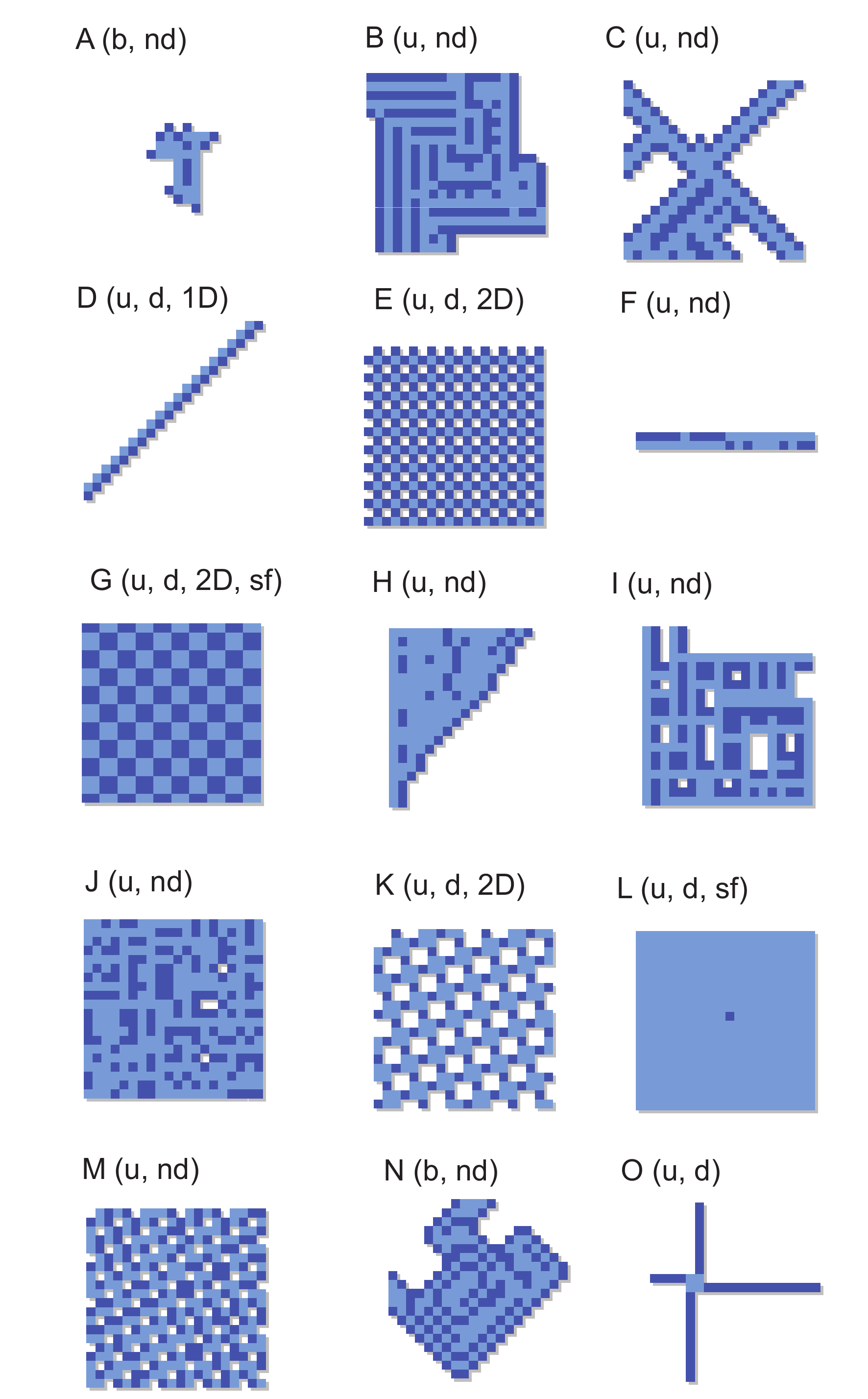}
\caption{ \footnotesize (colour online) Illustration of some UND polyomino types resulting from growth of genomes with $n_t = 2$, $n_c = 8$. UND polyominoes form the majority of achievable structures. The two colours label the two different tile types that may be involved in assembly. Letters in brackets denote whether the structures are bound (b), unbound (u), deterministic (d), non-deterministic (nd), space-filling (sf) and periodic in one (1D) or two (2D) dimensions.}
\label{unnon}
\end{center}
\end{figure}

\begin{figure}
\begin{center}
\includegraphics[width=6.3cm]{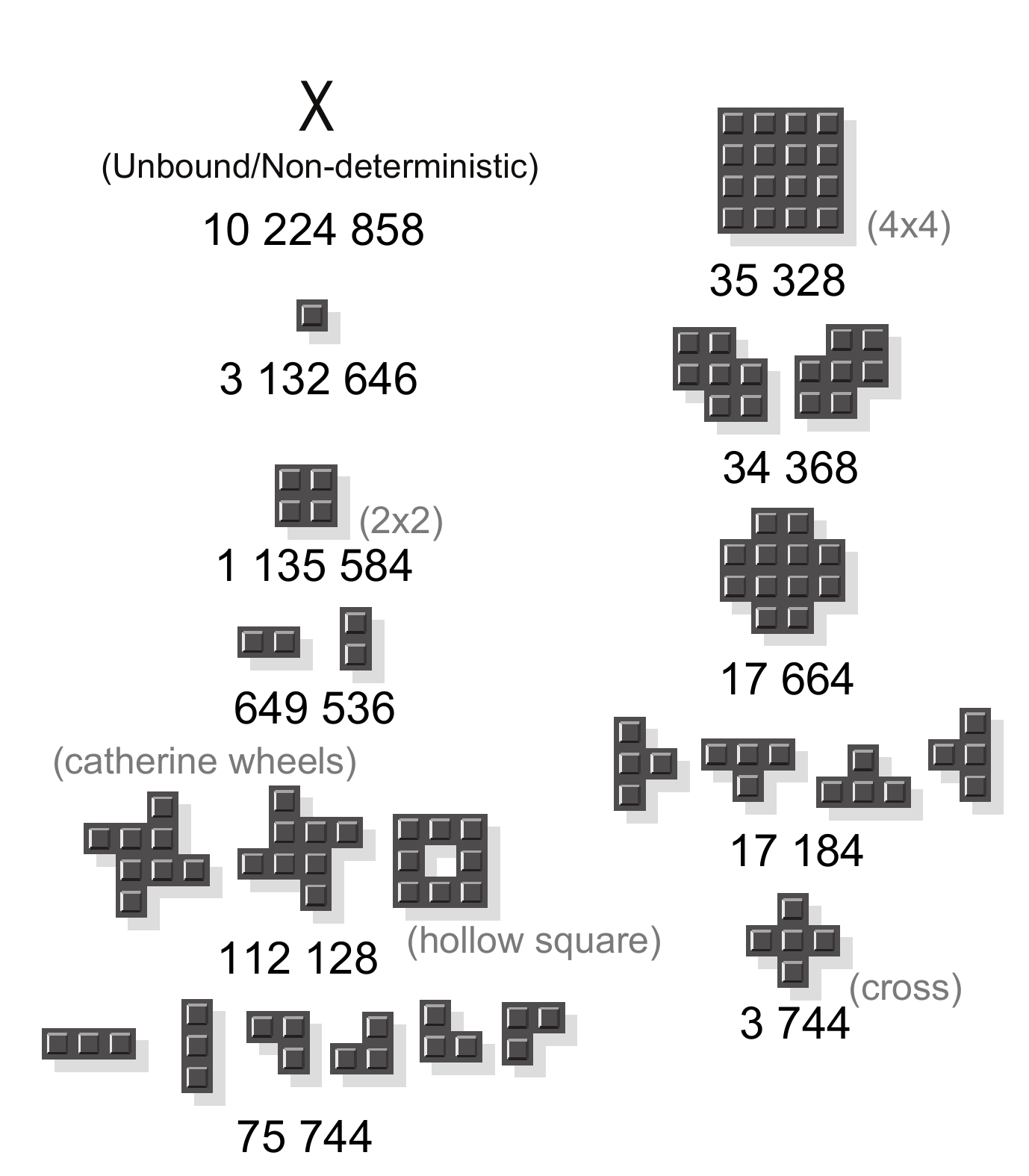}
\caption{ \footnotesize  $61.1\%$ of the polyominoes   that can be grown from genomes with $n_t = 2, n_c = 8$ are UND structures (X).  The rest are  bound, deterministic polyominoes. The number of genomes that encode for each polyomino are given -- for sets of structures with identical values, each structure occurs the given number of times in genome space. Some structures are given names for ease of reference in the text.}
\label{23polys}
\end{center}
\end{figure}

Rule sets in our model may result in \emph{unbound} structures: those where the assembly process proceeds in at least one direction without termination. Unbound structures may result, for example, from a set of one or more tiles that bonds to itself repeatedly, forming an endless chain of repeated units, as illustrated in Fig.~\ref{undesir}(a).

Self-assembly in biology may also yield unbound structures. Proteinaceous structures consisting of extended sets of repeated units include helical protein filaments such as microtubules \cite{amos2004microtubule}, actin filaments \cite{reisler2007actin} and tobacco mosaic virus \cite{kegel2006physical}; two-dimensional arrays such as S-layers \cite{sara2000s} and purple membranes \cite{krebs2000structural}; and even three-dimensional crystals \cite{doye2006protein}, although some biological mechanism must usually be present to regulate the size of these assemblies and prevent them being truly unbound \cite{goodsell2000s}.

Another assembly feature that may result from our model is \emph{non-determinism}, whereby the same set of rules may lead to different structures forming in the growth phase. This non-determinism is due to the inherent stochasticity in the assembly process. Non-determinism may arise when a tile edge is capable of bonding to more than one other tile edge, which may occur, for example, when the partner to a given edge type appears on more than one tile within the rule set, as in Fig.~\ref{undesir}(b). 

Non-determinism may occur in several ways. \emph{Shape non-determinism} is the least subtle form, whereby the overall shape of the produced structure (disregarding any detail of tile types, sides and orientations) differs stochastically in different assembly runs (see Fig. \ref{undesir} (b) for example). \emph{Tile non-determinism} occurs when the same overall structure is achieved for all runs, but sites within the structure are occupied by different tile types stochastically. \emph{Orientational non-determinism} occurs when the structure is both shape- and tile-deterministic, but tiles within the structure differing stochastically in orientation between assembly runs (an example of this is the structure in Fig. \ref{illus}). Another type of non-determinism, \emph{steric non-determinism}, may also occur as a result of the different speeds of growth in two directions that converge on the same point: if two arms of a structure pass through the same lattice point, the structure will differ depending on which arm arrives there first and hinders growth of the other. This type of non-determinism does not require the multiple bonding edges mentioned above, and is thus hard to detect through observation of the genome. 

In biology, non-determinism can also occur in a number of ways. Some closely related proteins coassemble into complexes of well-defined size and shape, but in which the identity of the protein at any position is random. An example of this phenomenon is in the seeds of pea plants \cite{stoger2001pea} where the legumin protein is formed by a number of paralogous genes, which result in hexamers containing randomly assorted subunits of similar but distinct polypeptide sequences.  This example would correspond to tile non-determinism in our model. There also exist examples of shape non-determinism in biology, where proteins, such as certain heat shock proteins, assemble into clusters with a polydisperse distribution of sizes \cite{laganowsky2010crystal, benesch2010quaternary}.

Finally, our assembly model may yield structures that are \emph{bound} (of finite size) and \emph{deterministic} (in which the self-assembly process always forms the same structure, with a specific shape). The majority of protein quaternary structures fall into this category \cite{goodsell2000s}. 

For completeness, we note that there is incomplete overlap between the sets of non-deterministic and unbound structures: rule sets may code for outputs that are unbound but deterministic, bound but non-deterministic, bound and deterministic or unbound and non-deterministic. In this study, we will focus on bound, deterministic structures, and will refer to structures not meeting these criteria as UND structures (unbound or non-deterministic). Some examples of these structures are shown in Fig. \ref{unnon}.

All these forms of non-determinism can in theory be detected by running each growth phase a large number $k$ times and comparing the output each time.  We shall employ $k = 10$, a value that was confirmed through preliminary investigation to detect most non-deterministic structures while retaining computational speed.
 In this investigation, we choose tile and orientational determinism as our desirable criterion.

\subsection{\label{secga}Genetic Algorithm Details}
Genetic algorithms (GAs) are a class of optimisation procedures that employ operators based on evolutionary biology to reach a solution to some problem \cite{goldberg1989genetic, mitchell1998introduction}. Typically, GAs involve a \emph{population} of $N$ individuals, each representing a trial solution to a problem. A \emph{fitness function} quantitatively measures the performance of an individual at solving the required problem. 

GAs take place over a number of time steps or \emph{generations}. Each generation, the fitness function is used to assign a fitness value $f_i$ to each genome $i$ in a population of $N$ individuals. A selection operator is then applied, selecting genomes for reproduction according to their fitness values, with high-fitness genomes being preferentially selected. The $N$ rule sets comprising the next generation are then formed from selected genomes. We employ the roulette-wheel selection method~\cite{goldberg1991comparative}, where the probability $P(i)$ of a genome $i$ being selected is proportional to its fitness: $P(i) = f_i/\sum_j f_j$. 


A common practise in the implementation of GAs is to preserve a certain number of the fittest individuals in a population from one generation to the next. This approach is termed \emph{elitism}, with a proportion $\epsilon$ of fit individuals preserved, immune to the effects of mutation \cite{mitchell1998introduction}. We will explore the use of elitism in Section \ref{secelite} but will generally set $\epsilon = 0$.
 

GAs may employ \emph{crossover}, modelling recombination. Without crossover, in the asexual regime, new individuals begin as cloned copies of selected genomes. With crossover, modelling sexual reproduction, new individuals are formed by selecting two `parent' rule sets from the old generation, forming a new rule set by combining the rule sets of these parents. The crossover scheme we employ is single-point crossover, where the first $L_R$ bits from one parent and the last $L-L_R$ bits from the other are combined to form a new individual, and $L_R$ is chosen randomly from $[0,L]$.

The implementation of crossover in a simulation is controlled by the crossover rate $R$, giving the proportion of new genomes that are formed through crossover. For simplicity, we will only employ values of $R = 0$ (corresponding to asexual reproduction) and $R = 1$ (corresponding to sexual reproduction).

Another genetic operator used in GAs is \emph{mutation}, whereby individuals in a generation undergo stochastic changes to their rule sets. We employ \emph{point mutation}, whereby each bit in the genome is flipped with probability $\mu$.

Genomes may contain redundant information, with a tile type being coded for more than once in the binary string. In addition, information on tiles and edges that do not play a role in the assembly of the final structure may be included in the genome. This unused information in genomes allows neutral mutation to progress. A genome may also, in the aforementioned non-deterministic case, code for many different polyomino structures, and the same structure may be produced by more than one genome, providing a many-to-many mapping.

\section{\label{secland}Search Space Analysis}

\begin{figure*}
\begin{center}
\includegraphics[width=18cm]{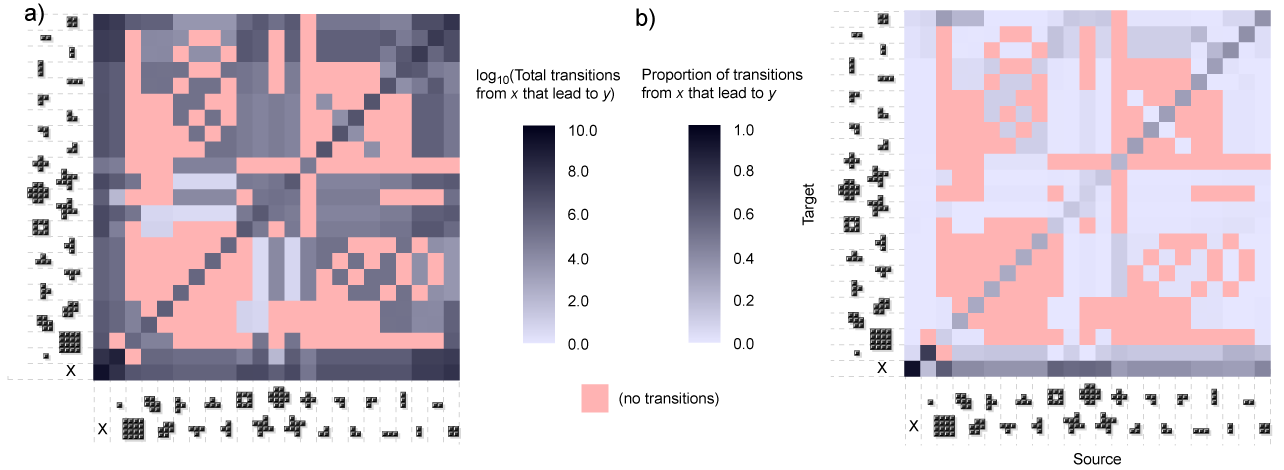}
\caption{ \footnotesize (colour online) Transitions between polyominoes in the $\mathcal{S}_{2,8}\,(n_t = 2, n_c = 8)$ system. (a) The value of a pixel denotes the total number of single-point mutations that result in a change from phenotype $x$ to phenotype $y$, over all genotypes in $\mathcal{S}_{2,8}$ that encode $x$ and $y$. (b) The value of a pixel denotes the average proportion of mutations that cause an $x \rightarrow y$ transition, where the average is taken over all occurrences of $x$ in $\mathcal{S}_{2,8}$. Pink pixels denote transitions between phenotypes that cannot be accomplished with a single mutation.}
\label{land}
\end{center}
\end{figure*}

The process of evolution can be viewed as an optimisation process on the high-dimensional search space defined by all possible genomes \cite{wrig32, maynardsmith1970protein}. An advantage of our self-assembly model is that the search space for simple structures can be fully characterised. We first investigate the structure of the search space for a polyomino model with two tiles ($n_t = 2$) and up to eight colours ($n_c = 8$), allowing three bond types ($1\leftrightarrow 2, 3\leftrightarrow 4, 5\leftrightarrow 6$), with colours 0 and 7 corresponding to neutral edge types. Each of the 8 tile edges can be represented by $\log_2 8 = 3$ bits, giving a binary genome of length $L = 24$. The search space therefore consists of $2^{24} \simeq 1.6 \times 10^7$ individuals. We will refer to search spaces as $\mathcal{S}_{n_t,n_c}$, labelled by the number of blocks (tiles) $n_t$ and number of colours $n_c$, so that the aforementioned search space is $\mathcal{S}_{2,8}$.

We adopt the convention that the first tile encoded in the genome is the assembly nucleus, and its initial orientation is specified by the order in which its edges are encoded, with the top edge first and others following in a clockwise direction. We then exhaustively evalulate all polyomino structures that may be constructed in this system. The majority are UND structures, some examples of which are shown in Fig.~\ref{unnon} to illustrate the diversity of achievable forms. These structures include non-deterministic, bound structures (for example, A in Fig.~\ref{unnon}), deterministic structures that are translationally periodic in one (D) or two dimensions (E, K), some of which may be space-filling (G). Unbound structures displaying shape- but not tile-determinism order also exist (F, M).

The resulting structures are illustrated, along with the volume of search space they occupy, in Fig.~\ref{23polys}. Sets of genomes encoding the same phenotype form the \emph{neutral network} of a given phenotype. The large differences in neutral network size corresponding to different phenotypes are related to the differing amounts of information required to specify bonds for different phenotypes. For example, the single tile phenotype only requires an absence of any bonding edges rather than any specific interaction pairs, and correspondingly occupies a large volume of genome space. By contrast, the $4 \times 4$ block phenotype requires two interacting pairs of edges, at specific positions relative to each other, and the number of genomes fulfilling these criteria is much smaller. 

In addition, all single mutation transitions were recorded, identifying the effect of every possible single mutation on every possible genome --- which may change the phenotype or be neutral (with no phenotypic effect). Fig.~\ref{land}(a) depicts the number of possible single-mutation transitions between different phenotypes, whilst Fig.~\ref{land}(b) depicts the probability of a transition to another structure given an initial structure. In (b), the total number of transitions between two phenotypes are normalised by the number of genomes encoding the $x$-axis phenotype (see Fig. \ref{23polys}). The resulting quantity measures the average number of mutations in a genome that encodes phenotype $x$ that cause a transition to phenotype $y$. 

The Fiedler eigenvalue method \cite{massen2006thermodynamics} was used to arrange the phenotypes in Fig.~\ref{land} to maximise the ``blockiness'' of the resulting matrix by clustering rows and columns whose elements follow similar trends. This method noticeably groups modularly-related polyominoes --- for example, the $2 \times 1$ and $3 \times 1$ structures are clustered together, as several single-mutation changes allow transitions between these structures through the addition or subtraction of a single block. This clustering reflects the fact that pairs of polyominoes that share modules (tiles or bond sequences) are more closely connected in genome space than unrelated structures.

Fig.~\ref{land}(b) shows that the majority of single mutations from a given phenotype are either \emph{neutral}, preserving the phenotype -- leading to high diagonal values in the plot -- or cause a transition to a UND or single-tile phenotype. The fraction of neutral mutations is noticeably smaller for larger polyominoes (for example, the `catherine wheel' structures and the $4 \times 4$ block have diagonal values under 0.3) than smaller ones (for example, the single tile, $2 \times 1$ blocks and the $2 \times 2$ block have diagonal values over 0.6), partially because genomes encoding small polyominoes contain more redundant information than those encoding large polyominoes.

Another observable feature of the search space is that, for several phenotypes, the most common result of mutations that are not neutral and do not result in a UND phenotype is a loss of part of the structure associated with the phenotype. For example, a significant proportion of mutations lead from the $4 \times 4$ block to the $2 \times 2$ block, removing the outer `shell' of tiles. The T-shaped tetrominoes also show many transitions to the L-shaped triominoes, as one tile is lost from the phenotype. These triominoes in turn show many transitions to the $2 \times 1$ blocks, from the loss of another bonded tile. 

Fig.~\ref{land}(b) gives a measure of the average \emph{robustness} and \emph{evolvability} \cite{wagner2005robustness} of a given phenotype. The diagonal values give the phenotypic robustness, measuring the average (over all genomes that encode a given phenotype) number of possible mutations that preserve phenotype. This averaging gives a mean phenotype robustness rather than the robustness value for any individual genome \cite{wagner2008robustness}. Phenotypic evolvability can be measured in two different ways. Firstly, a sum over off-diagonal values gives the number of mutations that result in a useful (non-UND) phenotypic change.   Secondly, the number of non-zero off-diagonal values in a column give the number of different phenotypes that can be accessed from the source phenotype.  The first measure can be used to describe the probability that a non-neutral mutation will result in a useful phenotype.  The second is more closely related to Wagner's definition of phenotype evolvability \cite{wagner2008robustness}: it measures the diversity of phenotypes accessible from the neutral network of a given phenotype.    In our model, robustness and evolvability are related differently in different phenotypes: the catherine wheel structures are highly evolvable according to both the above definitions, but have low robustness (about 0.3), whereas the $2 \times 2$ square has high evolvability and high robustness (about 0.6).





\section{\label{secevdyn}Evolutionary Dynamics}

\subsection{\label{secex}Evolving Polyomino Size}

\begin{figure}
\begin{center}
\includegraphics[width=8cm]{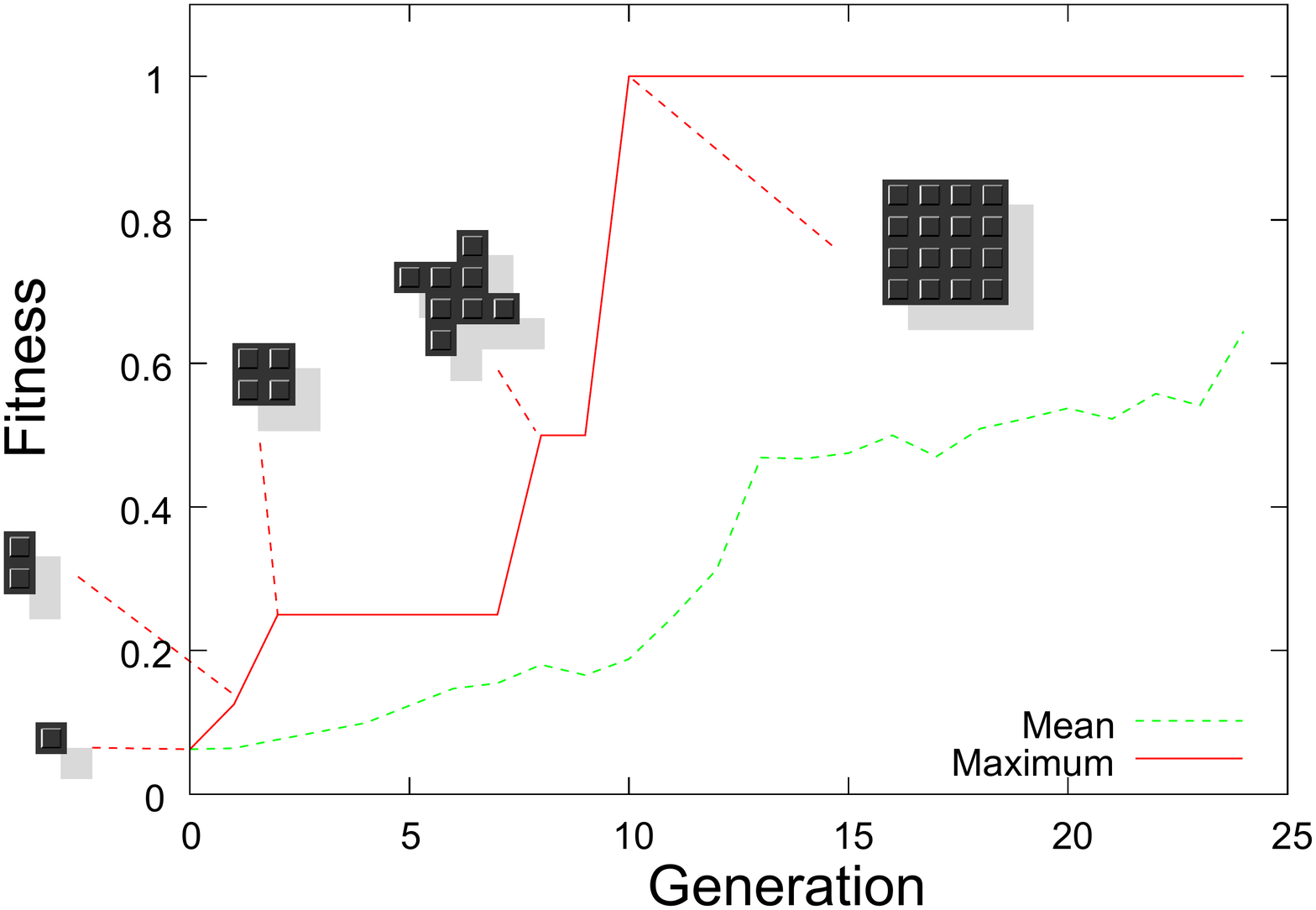}
\caption{ \footnotesize (colour online) Fitness curves during a typical evolutionary run. A population of genomes is evolved towards a structure of size $s \geq 16$ using the fitness function in Eqn. \ref{sizeff}, at $\mu L = 0.5$, $R = 0$, $N = 10$. The plot shows the mean (dashed) and maximal (solid) fitness within a population as time progresses.}
\label{B2fitcurve}
\end{center}
\end{figure}

In evolution, selection drives a system towards high-fitness phenotypes (analogous to a thermodynamic drive towards low-energy structures), and entropic effects favour those structures that occupy a large proportion of search space. This interplay of fitness and entropic terms is analogous to the concept of free energy in thermodynamics, and indeed several studies have analysed evolution using a `free fitness' quantity \cite{iwasa1988free, sella2005application}. It may be expected that the importance of a given phenotype in evolutionary dynamics is related to several factors, including the fitness of the phenotype and how frequently genomes that produce it occur in the search space. For example, if fitness is defined as proportional to polyomino size, we may expect large structures that occupy a large volume of search space (i.e.\ with relatively large neutral networks)--- like the `catherine wheels' and the $4 \times 4$ block in Fig.~\ref{23polys} --- to play important roles in evolutionary pathways.

Having characterised the $\mathcal{S}_{2,8}$ search space in detail, we now proceed to simulate evolution on a fitness landscape in this search space, with a particular aim being to relate the evolutionary dynamics back to the structure of the underlying search space. We use a specific fitness function to drive evolution towards a given target with a GA. A simple example is evolution towards a bound, deterministic polyomino matching or exceeding a certain size, using the fitness function:
      
\begin{equation}
F(P_1, P_2, ..., P_k; s^*) = \left\{ \begin{array}{ll}
& 1, \\
& \hspace{0.25in}s(P_1) \geq s^*\, \mbox{and all}\,P_i\\
& \hspace{0.25in}\mbox{identical and bound;} \\
& s(P_1) / s^*, \\
& \hspace{0.25in}s(P_1) < s^*\,\mbox{and all}\,P_i\\
& \hspace{0.25in}\mbox{identical;} \\
& 0, \\
& \hspace{0.25in}\mbox{$P_i$ unbound or} P_i \not= P_j \\
& \hspace{0.25in}\mbox{for any $i, j$.} \end{array} \right.
\label{sizeff}
\end{equation}


Here the fitness function takes a set of polyominoes $\{P_1, ..., P_c\}$ produced through $k$ repeats of the assembly process, and a desired size $s^*$. The function returns a zero fitness value if the set of polyominoes is UND, and a fitness value proportional to polyomino size for bound, deterministic structures. A value of one means that a solution matching the size criterion has been found.

We note that the previous section suggests that only a small minority of the possible mutations to a genotype lead to a phenotype of larger size. However, it may be expected that on the rare occasions that such mutations do take place, selection will allow these phenotypic changes to be retained and propagate through the population.

Fig.~\ref{B2fitcurve} shows the time evolution of a population of polyominoes towards the target $s^* = 16$. On the $\mathcal{S}_{2,8}$ landscape, only one phenotype fulfills this criterion: the $4 \times 4$ block. We employ what we will term \emph{zero initial conditions}, in which every bit in every genome at the start of the simulation is set to zero. In the self-assembly implementation described above, this approach means every initial genome encodes a single tile phenotype, which is laid down and incapable of further bonding. 

The simulation begins with the trivial, single-tile phenotype, then quickly `discovers' beneficial interactions, increasing the size of the largest phenotype in the population first to two then to four, with the $2 \times 2$ square structure being discovered.  The mean fitness lags behind the maximal fitness, as many members of the population will still possess lower fitness values -- the mean fitness rises only gradually above the value corresponding to the single tile phenotype, as the information for the $2 \times 2$ square structure does not immediately propagate through the whole population. The slow spread of information is due to both the finite fitness advantage resulting from the larger size of the square structure, and the possibility of further mutations leading to UND structures. After several generations, a further beneficial interaction is discovered, creating the `catherine wheel' octomino, and in the next generation this structure is expanded upon to form the $4 \times 4$ structure. Note that the catherine wheel structure is one of only a few phenotypes exhibiting a single-mutation transition to the $4 \times 4$ block (see Fig. \ref{land}). 

The discovery of the $4 \times 4$ block leads to a sharp rise in the mean fitness, which lasts several generations before flattening. This flattening is due to the non-zero mutation rate and the high transition probability between the $4 \times 4$ block and other structures of lower fitness.  This mutational entropy means that for a finite $\mu$, perfect adaptation is not reached in this system.    The simulation is terminated when more than half the population has maximal fitness.

\subsection{\label{secmur}Varying Mutation Rate}

\begin{figure}
\begin{center}
\includegraphics[width=8cm]{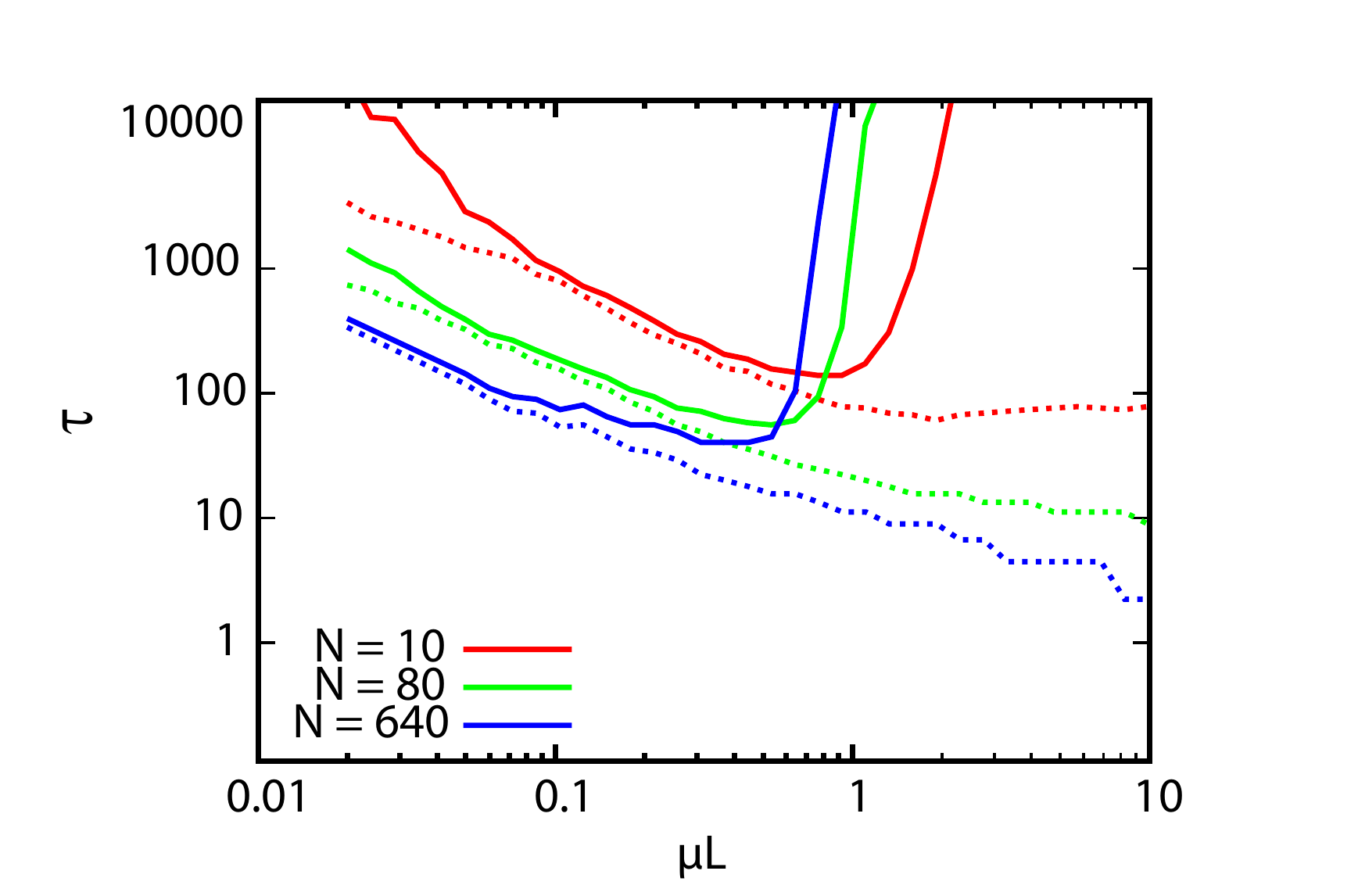}
\caption{ \footnotesize (colour online) Adaptation time $\tau$ (solid lines) and discovery time $\tau_D$ (dashed lines) in generations, in $\mathcal{S}_{2,8}$ evolving to $s^* \geq 16$, with mutation rate $\mu$, at different $N$ and with $R = 0$.}
\label{aplot}
\end{center}
\end{figure}

\begin{figure}
\begin{center}
\includegraphics[width=8cm]{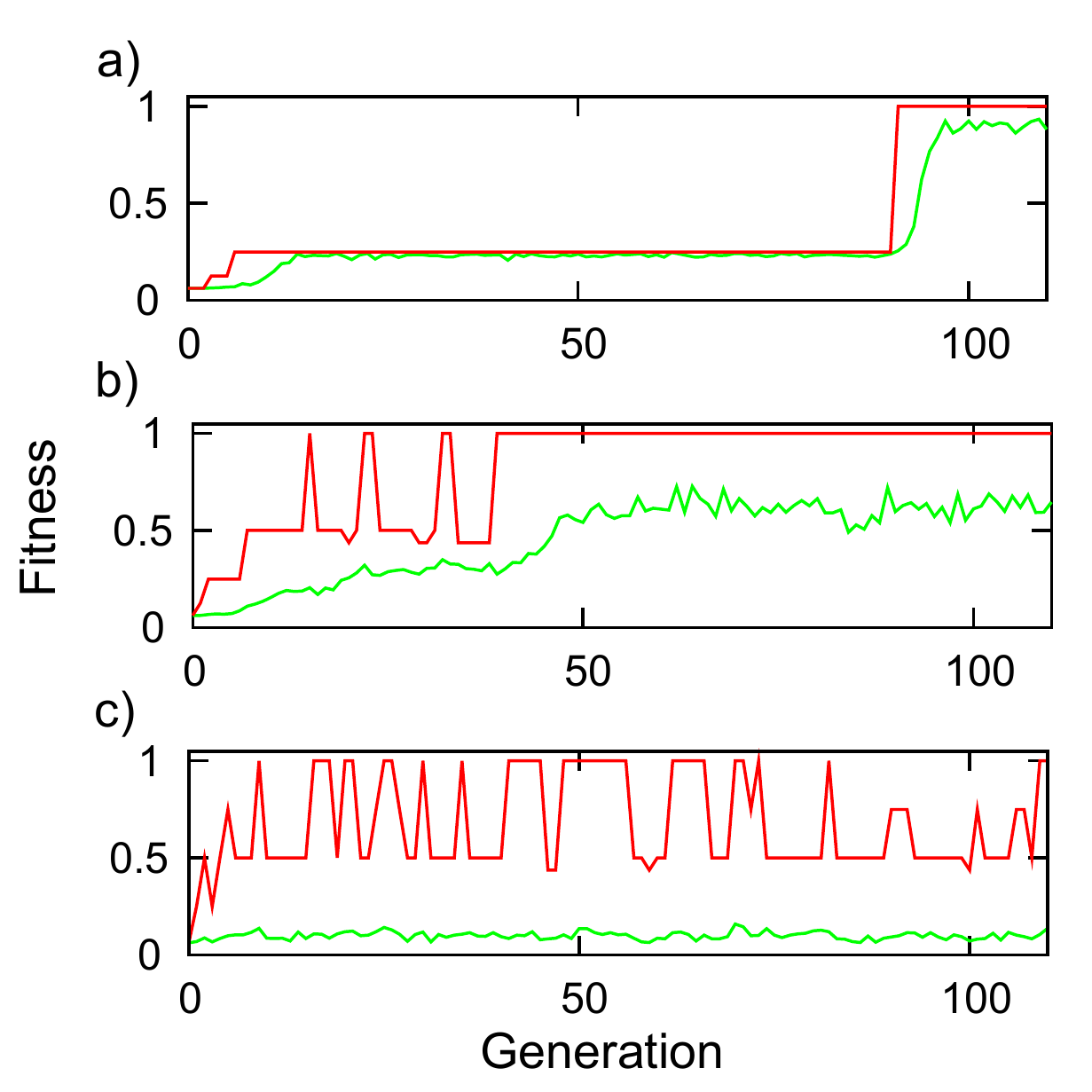}
\caption{\footnotesize (colour online) Fitness curves with time for simulations in $\mathcal{S}_{2,8}$ evolving to $s^* \geq 16$, with $N = 80, R = 0$ and (a) $\mu L = 0.1$, (b) $\mu L = 0.5$, (c) $\mu L = 2$. Maximum fitness in the population is shown in red (upper curves) and mean fitness is shown in green (lower curves). Adaptation, defined as the point where $50\%$ or more of the population has maximal fitness, occurred at generations 95 for (a) and 51 for (b). (c) failed to adapt within the $20\,000$ generation cutoff.}
\label{fitcurves}
\end{center}
\end{figure}

In GA experiments, the \emph{discovery time} $\tau_D$ measures how long a system takes to produce a single copy of a maximally fit solution, giving an indication of the speed at which evolution progresses. Specifically, $\tau_D$ is the first generation in which at least one genome encoding a maximally fit solution is present. 

The distribution of $\tau_D$ in an ensemble of GA experiments is generally observed to be long-tailed, with infrequent occurrences of very high discovery times.  Due to computational limitations, we generally employ a cutoff of $20\,000$ generations in our GA runs. As these rare, high values can skew the mean of such a distribution, we use the median of the distribution as a measure for $\tau_D$, as this statistic is less prone to skew from the rare events and artefacts from the imposed cutoff. $1\,000$ GA runs were performed for each data point in the following plots.

We measured the value of $\tau_D$ in the $\mathcal{S}_{2,8}$ system, as a function of mutation rate $\mu$ at a range of population sizes $N$. We set $R = 0$ and use zero initial conditions. Fig.~\ref{aplot} shows the results. $\tau_D$ decreases monotonically with $\mu$ except in the case of low $N$, where a slight increase at high $\mu$ is observed. The decrease in $\tau_D$ at high $\mu$ is due to the allowed larger steps across search space and a more explorative search. The slight increase in $\tau_D$ at high $\mu$ in the low $N$ case may be due to the inability of completely random search to efficiently explore the search space with a small population -- in other words, either some memory of previously discovered information or a large population is required for optimal search. We will see in Section \ref{seclarge} that the monotonic decrease in $\tau_D$ for larger population sizes is due to the small size of the $\mathcal{S}_{2,8}$ search space, and that $\tau_D$ exhibits an optimum with $\mu$ in larger search spaces.

Another timescale of interest in evolutionary simulations is the \emph{adaptation time} $\tau$ of a system, measuring how long a solution, designated as maximally fit, takes to dominate the population. We measure this quantity as the first generation in which more than half the population has maximal fitness. The reason this criterion was chosen is that, due to 
the high proportion of deleterious mutations that decrease fitness (see Section \ref{secland}), full adaptation is unlikely to occur in reasonably sized populations at finite $\mu$ due to the likelihood of at least one phenotype-changing mutation occurring in a population.

Fig.~\ref{aplot} shows $\tau$ values for the $\mathcal{S}_{2,8}$ system, with $R = 0$. A general observation is the presence of an optimal mutation rate $\mu^*$, at which $\tau$ is a minimum.  The optimal mutation rate arises from the following competition. At  very low $\mu$, $\tau$ increases divergently as $\mu$ decreases. This increase in $\tau$ at low $\mu$ is steeper at low $N$ than at high $N$.   The reason is simply that at low mutation rates, it takes a long time for the system to discover new phenotypes, and this is made worse in smaller populations.     On the other hand,  arguments from population genetics \cite{nowak2006evolutionary} suggest that full adaptation of a population becomes increasingly difficult  for $\mu \gtrsim 1/L$, due to mutational entropy.  Thus one expects an optimal mutation rate for adaptation around $\mu \approx 1/L$.

Fig.~\ref{fitcurves} shows examples of the time evolution of the fitness during simulations at a range of $\mu$ values (low: $\mu L = 0.1$, intermediate: $\mu L = 0.5$, high: $\mu L = 2$). At low $\mu$, the mean fitness closely tracks the maximal fitness, as diversity is low and the population is confined around a small region of genome space. The behaviour is due to the high correlation between generations: as little change is introduced to the gene pool through mutation, diversity in the population is low.  

At high $\mu$, the mean fitness fluctuates around a low value, dominated by the entropic drive towards common, low-fitness structures (as most mutations are deleterious --- see Fig.~\ref{land}). In this regime, the population is decorrelated, and highly genetically diverse --- resembling a random search across genome space. 

Behaviour at $\mu L = 1$ is intermediate between these regimes, with some diversity resulting in a rather lower mean fitness than maximal fitness, but a clear relationship between the two showing that information is not being lost through decorrelation.
  
The relationship between mean and maximal fitness also depends on the robustness of the phenotypes within a population. In Fig. \ref{fitcurves} (b), the mean and maximal fitness values are closer in magnitude for a local optimum (around generation 40) than for the global optimum (generation 43 onwards). This difference suggests that the robustness of the global optimum is lower than that of the local optimum, as the population has more difficulty adapting to the fitter phenotype.

We will use the terms \emph{exploration} and \emph{exploitation} to refer to the two regimes observable at high and low $\mu$, respectively. Exploration refers to the random search regime at high $\mu$, where genome space is explored uniformly and randomly, and the entropic effect of mutation is too high for the population to become localised and adapt. Exploitation refers to the highly-correlated regime at low $\mu$, where evolution progresses through small changes made to existing information, resembling a ``hill-climbing'' process with a low diversity. The intermediate $\mu$ regime may be thought of as providing a combination of these two effects, with enough exploration to allow escape from local optima and enough exploitation to experience a drive to higher fitness values.

\subsection{\label{seclarge}Comparing Search Spaces}
To investigate the effect of changing the search space for the system, we next considered the $\mathcal{S}_{6,8}$ space, involving $n_t = 6$ blocks rather than the $n_t = 2$ used previously. Genome length is now $L = 72$, with $2^{72} \simeq 4.7 \times 10^{21}$ points in search space, more than 14 orders of magnitude larger than the $\mathcal{S}_{2,8}$ space. We used a sampling approach, investigating $10^8$ points in $\mathcal{S}_{6,8}$, to investigate how the structure of this new search space may affect the search for an $s \geq 16$ structure. Firstly, a larger number of genomes in the new space encode for such a structure, with many possible ways of achieving the $4 \times 4$ square and other, more diverse structures with $s \geq 16$. However, the associated exponential increase in the overall size of the search space means that a smaller \emph{proportion} of genomes encode a structures with $s \geq 16$, with many more genomes now producing small or UND polyominoes. 

\begin{figure}
\begin{center}
\includegraphics[width=8cm]{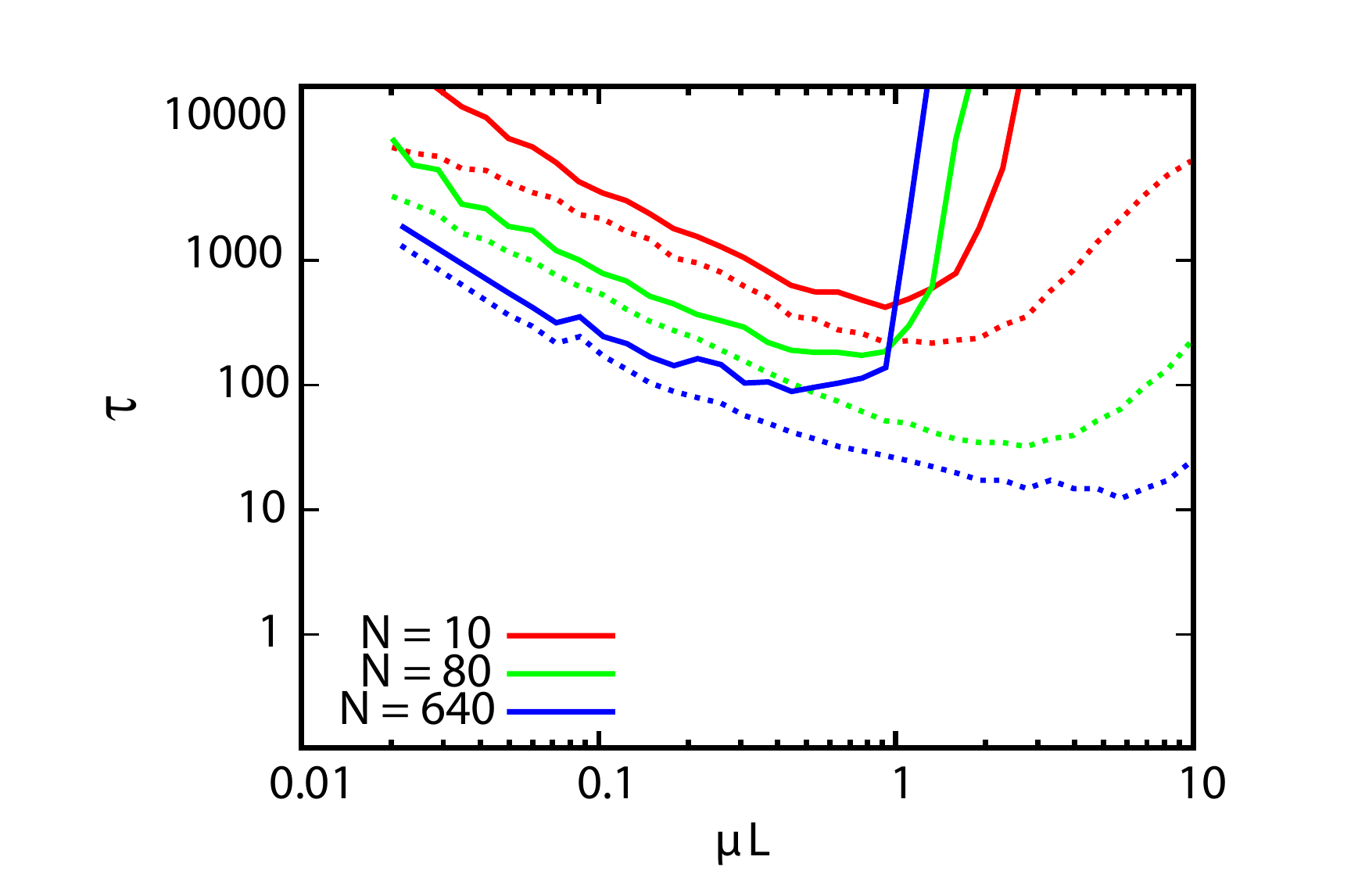}
\caption{ \footnotesize (colour online) Adaptation time $\tau$ (solid lines) and discovery time $\tau_D$ (dashed lines) in $\mathcal{S}_{6,8}$ evolving to $s^* \geq 16$, with mutation rate $\mu$, at different $N$ and with $R = 0$.}
\label{bplot}
\end{center}
\end{figure}

Fig.~\ref{bplot} shows the $\tau$ and $\tau_D$ behaviour with $\mu$ in $\mathcal{S}_{6,8}$.  In this plot, we see first of all that even though the search space is many orders of magnitude larger, the optimal adaptation and discovery times are at most an order of magnitude larger.  The qualitative behaviour of the discovery time $\tau_D$ also shows an important difference from the simpler $\mathcal{S}_{2,8}$ system. This measure now exhibits an optimum with $\mu$, generally around $\mu L  \geq 1$. At higher $\mu$ values, $\tau_D$ increases, indicating that the large steps performed by high-$\mu$ search in this case are not beneficial. This optimum arises from a tradeoff between exploration and exploitation: the system must have a high enough $\mu$ to successfully explore a range of genome space, but must have a low enough $\mu$ so that useful information is not lost. 

At high $\mu$, the gene pool decorrelates significantly from generation to generation, resulting in loss of information about intermediate-fitness structures that have been discovered. In the smaller $\mathcal{S}_{2,8}$ system, Fig.~\ref{aplot} suggests that this loss of information is not an important effect, as the highly random search afforded by high $\mu$ has a finite chance of discovering a suitable solution through exploration alone. However, in the exponentially-larger $\mathcal{S}_{6,8}$ space, random search has a very low probability of discovering a suitable solution, and exploitation of existing information is important in the discovery of better solutions.

\subsection{\label{secic}Initial Conditions}

\begin{figure}
\begin{center}
\includegraphics[width=8cm]{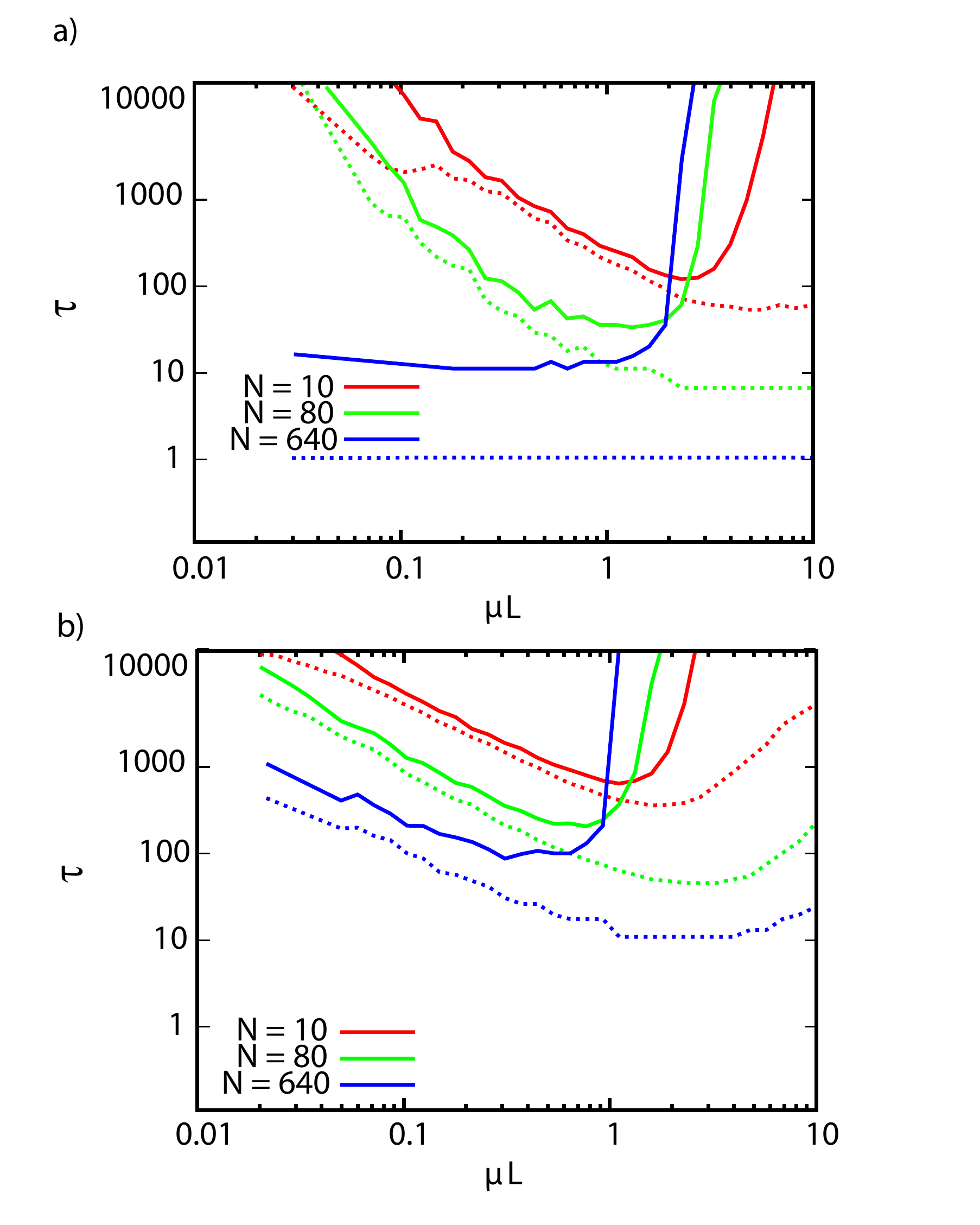}
\caption{ \footnotesize (colour online) Adaptation time $\tau$ (solid lines) and discovery time $\tau_D$ (dashed lines) with random initial conditions in (a) $\mathcal{S}_{2,8}$ and (b) $\mathcal{S}_{6,8}$, evolving to $s^* \geq 16$, with mutation rate $\mu$, at different $N$ and with $R = 0$.}
\label{deplot}
\end{center}
\end{figure}

Many studies of evolution employ \emph{random initial conditions}, where the initial population is randomised before numerical simulation \cite{kashtan2007varying, cohe05, oikonomou2006effects}. While this picture is appropriate for the modelling of randomly distributed alleles in a population, it is of dubious biological relevance when bits in a genome represent more fundamental units of genetic information, as it corresponds to an interbreeding population with entirely different, randomised genomes. In considering the evolution of a self-assembling system such as protein quaternary structure \cite{villar2009self, levy2008assembly}, it may be that the uniform population of trivial phenotypes afforded by our aforementioned zero initial conditions is more biologically relevant.

To compare the two scenarios, we ran simulations of the $\mathcal{S}_{2,8}$ and $\mathcal{S}_{6,8}$ systems with random, rather than zero, initial conditions. The results (Fig.~\ref{deplot}) show a significant departure from our results with zero initial conditions. The difference is particularly pronounced at high $N$, where the diversity provided by a large population of random genomes will lead to very low discovery times, as space can be explored very quickly from this start point before any adaptation takes place. In fact, the $N = 640$ $\mathcal{S}_{2,8}$ system shows a discovery time of one, as the proportion of search space corresponding to a solution ($35\,328 / 2^{24} \simeq 2.1 \times 10^{-3}$) is more than $1/N$ ($1/640 \simeq 1.6 \times 10^{-3}$), making it likely that at least one random genome in the initial population will already be a suitable solution. By contrast, this random search effect has little impact in the much larger search space of the $\mathcal{S}_{6,8}$ system.

\subsection{\label{secrec}Recombination}
We next set $R = 1$, modelling sexual reproduction. This parameterisation was observed to have little effect on the behaviour of $\tau$ values in the $\mathcal{S}_{2,8}$ and $\mathcal{S}_{6,8}$ systems with zero initial conditions, leading only to a slight increase in adaptation times for given $\mu$. The effect of setting $R = 1$ with random initial conditions was much more pronounced. In this case, discovery times were significantly reduced and adaptation times were raised in both systems, suggesting that recombination may act to increase the `effective mutation rate' experienced by a genome.

In this picture, recombination may act to decorrelate an offspring from both its parents if the genetic diversity in the population is high. This effect may be, to first order, absorbed into an effective mutation rate dependent on the diversity in the population. Random initial conditions ensure that this diversity is high, particularly for large $N$, and hence the steps across genome space caused by crossover may be large. This `genetic drift' acts in cohort with the bare mutation rate $\mu$, facilitating rapid discovery of solutions on the small $\mathcal{S}_{2,8}$ search space, but acting to hinder adaptation at higher $\mu$.
%
%

\begin{figure}
\begin{center}
\includegraphics[width=8cm]{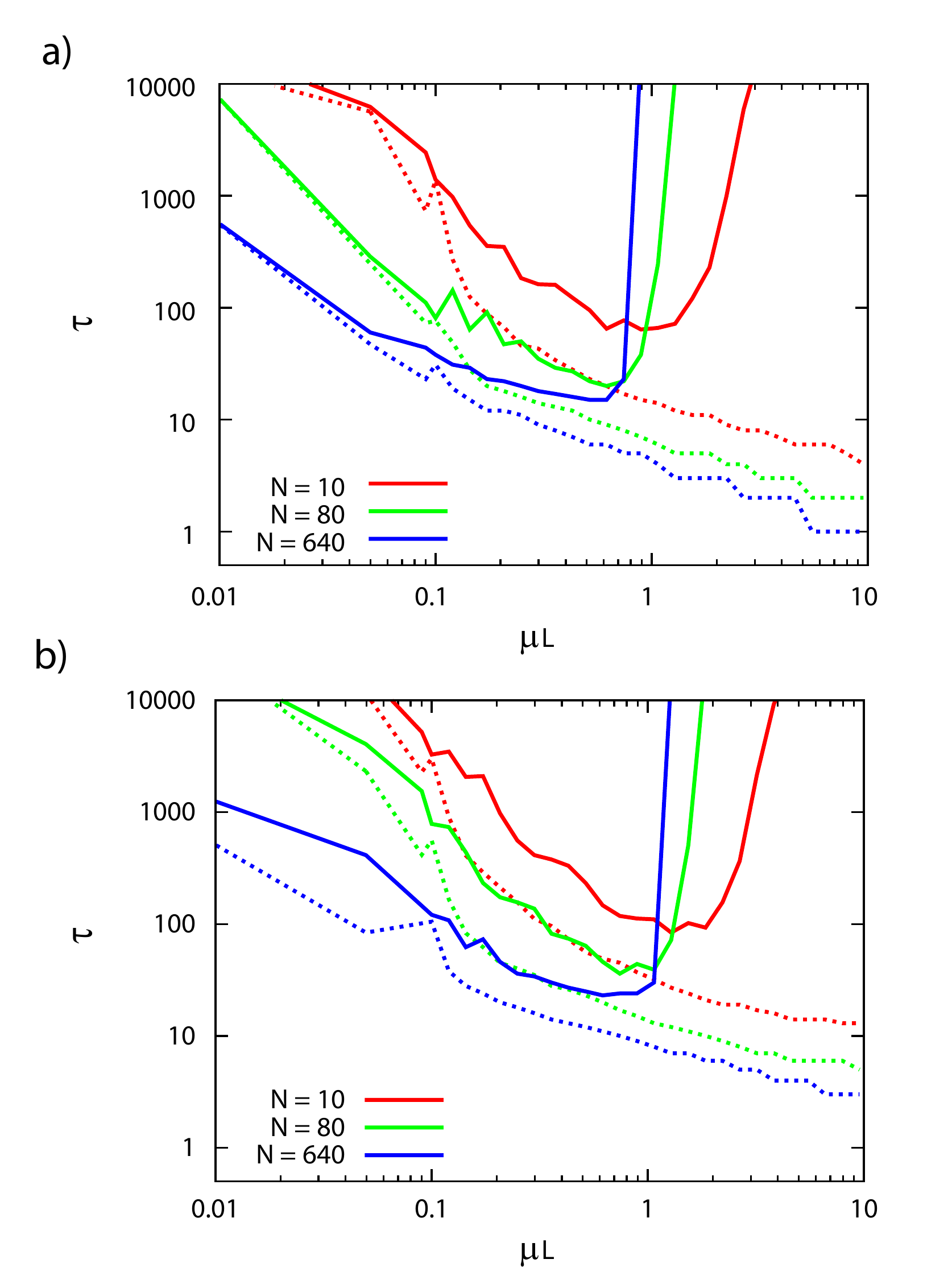}
\caption{\footnotesize (colour online) Adaptation time $\tau$ (solid lines) and discovery time $\tau_D$ (dashed lines) with $\epsilon = 0.1$ for (a) $\mathcal{S}_{2,8}$, (b) $\mathcal{S}_{6,8}$, with zero initial conditions. The increase in $\tau_D$ with high $\mu$ for $n_t = 6$ has vanished, and all $\tau$ values are lower than the $\epsilon = 0$ equivalents.}
\label{witheps}
\end{center}
\end{figure}

\subsection{\label{secelite}Elitism}

Optimisation-oriented applications of GAs often employ \emph{elitism}. In a population of $N$ individuals with elitism $\epsilon$ (where $\epsilon \in [0,1))$, the fittest $\epsilon N$ individuals are preserved totally intact from one generation to the next, immune to the action of mutation and recombination. In this way, the information within the fittest individuals --- the location of the highest peak thus far discovered --- is preserved, so that decorrelation from this point progresses more slowly and can never be complete. This approach is often beneficial for optimisation as it allows larger $\mu$ values to be used --- increasing exploration efficiency --- without loss of information about the current best solution.

The biological relevance of elitism is questionable. The problem arises from the immunity of the fittest individuals to mutation (and crossover, in a sexually reproducing population). This situation essentially corresponds to a number of extremely long-lived individuals which continually reproduce through their lifetimes, dying only when a fitter solution is found. 

Elitism can have a profound effect on the evolutionary dynamics of a model. Fig.~\ref{witheps} shows $(\mu, \tau)$ curves for a range of evolutionary scenarios with $\epsilon = 0.1$. These effects include a general reduction in $\tau$ values, showing that elitism is a useful tool in pure optimisation application of GAs. The increase in $\tau_D$ with high $\mu$ on $\mathcal{S}_{6,8}$ is no longer observed, as elitism retains information from one generation to the next, meaning that the search never becomes fully random. In experiments with recombination (not pictured), elitism also acts to stabilise the population, with adaptation observed in $\epsilon = 0.1$ simulations in some regimes that struggled to adapt with $\epsilon = 0$. 

\section{Homomeric Protein Tetramers}
\label{seclevy}

\begin{figure}
\begin{center}
\includegraphics[width=4cm]{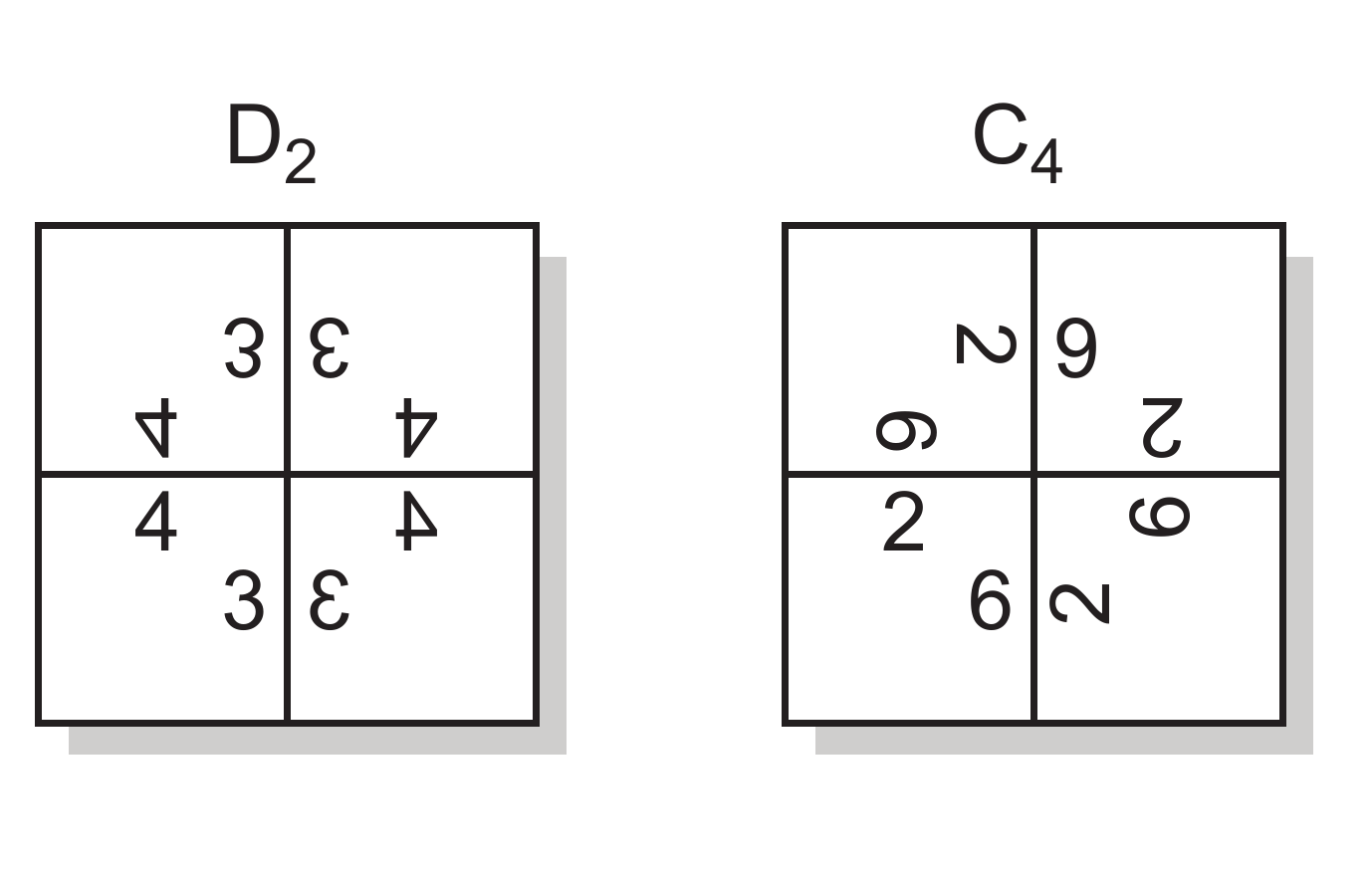}
\caption{Illustration of $D_2$ and $C_4$ symmetries in homomeric tetramers.}
\label{d2c4}
\end{center}
\end{figure}

It has been estimated that between $50$ and $70\%$ of proteins  form homomeric clusters \emph{in vivo}~\cite{levy20063d}. These complexes are usually symmetrical,  with each protein in an identical environment.  Homomeric tetramers, for example, may display cyclic symmetry ($C_4$) or dihedral symmetry ($D_2$).    The $C_4$ geometry involves only one type of interaction, whereas the $D_2$ complex involves at least two self-complementary interactions.   In an important recent study by Levy \emph{et al.} \cite{levy2008assembly}, it was shown that dihedral complexes are over 10 times more abundant than cyclic complexes with the same number of subunits.  Moreover, these authors found that the evolutionarily older interactions are typically stronger than the more recently evolved patches, and that the clusters dissassembled in a hierarchical fashion, with the newer (and weaker) bonds breaking first.     

The relationship between the strength of the patches and their evolutionary history, as well as the observed hierarchical dissassembly can be rationalized with simple statistical mechanical models~\cite{villar2009self}.     Similarly, the preference for dihedral over cyclic symmetry has been linked to the fact that for  $D_2$ structures, two pairs of identical edges bond (requiring self-complementary interactions or \emph{homointeractions}), whereas in $C_4$ structures, one pair of different edges bond (using non-self-complementary interactions or \emph{heterointeractions}).  Statistical models of the formation of homointeractions and heterointeractions have shown that the former have a wider distribution of energies than the latter.  It has been suggested that this wide distribution makes stable low-energy bonds easier to evolve using homointeractions than heterointeractions  which may result in a biological preference for $D_2$ structures~ \cite{lukatsky2006statistically,andre2008emergence,schulz2010dominance}.  Another reason for the preference for $D_2$ may be that evolution does not need to proceed to a tetramer structure in a single step, but can go through a dimeric intermediate.
 
 Our simple polyomino model cannot be used in its current form to study the strength of patches, and by extension, the hierarchical assembly/disassembly.  However, it can be used to investigate the effect of homo/heterointeractions and evolutionary intermediates on the evolutionary preference for $D_2$ over $C_4$.  In order to model this system, we must generalise our model to allow tetrameric structures to form in both symmetry configurations, as shown in Fig. \ref{d2c4}.  To do this, we allow building block tiles to `flip', so that, for example, tiles $\{1, 2, 3, 4\}$ and $\{1, 4, 3, 2\}$ are equivalent. The sides of building blocks now correspond to free, rather than fixed, necklaces \cite{necklace}. This condition reflects the fact that homointeraction interfaces require a rotation by $\pi$ radians with respect to each other to form a bond.

We first investigate the case where heterointeractions are equally easy to evolve as homointeractions. To achieve this, we choose a new interaction matrix such that the bonding pairs are: $3 \leftrightarrow 3, 4 \leftrightarrow 4, 2 \leftrightarrow 6$, with all other colours neutral. This setup was chosen so that, given zero initial conditions, the formation of two self-interacting edges involves the same number of mutations as the formation of a non-self-interacting bonding pair. Specifically, the discovery of colours 3 and 4 ($011$ and $100$) or 2 and 6 ($010$ and $110$) are equally likely, each requiring three beneficial mutations. We label this new search space $\mathcal{S'}_{18}$, with a characteristic number of self-interactions $n_{si} = 2$. We note that, given that $n_t = 1$ for this system, there is no distinction between the SFN and SGN cases mentioned in Section \ref{secmod}.   

\begin{figure}
\begin{center}
\includegraphics[width=8cm]{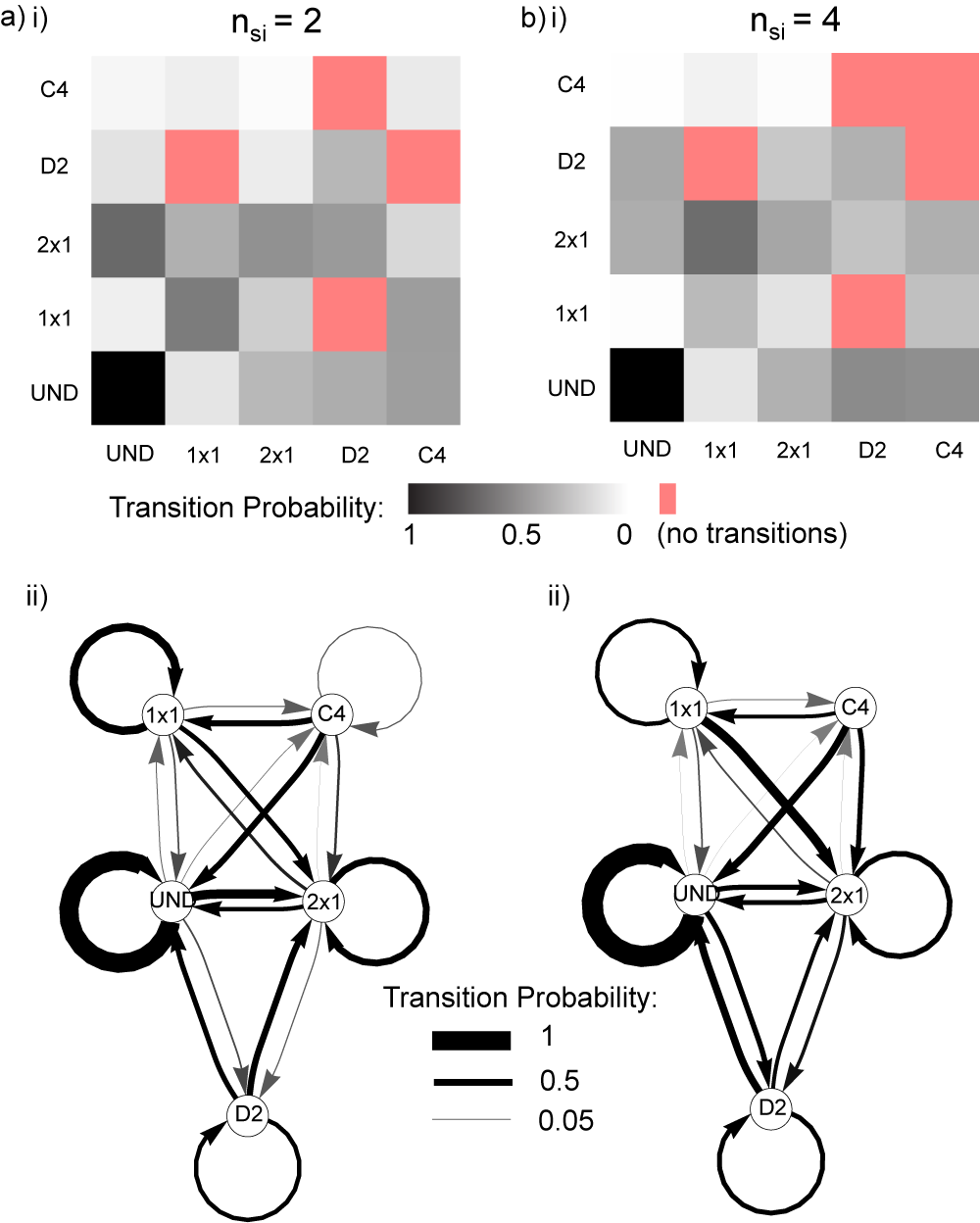}
\caption{(colour online) Transition probabilities for $\mathcal{S'}_{1,8}$. (a) Number of self-interacting colours $n_{si} = 2$ and (b) $n_{si} = 4$.  (i) Transition probabilities between phenotype $x$ and phenotype $y$. (ii) Transition probabilities represented in a network between phenotypes. The edge widths are proportional to their probability.  }
\label{s18}
\end{center}
\end{figure}

In a similar manner to that used for the $\mathcal{S}_{2,8}$ system in Section \ref{secland}, we can evaluate all possible structures in this new search space and the possible transitions between phenotypes (see Fig.~\ref{s18}).  There are $4\,096$ different possible genotypes, which are distributed among the possible phenotypes as shown in Table \ref{ngen}. A completely random search would thus display a $D_2$ structure frequency of 0.68. While the interactions are chosen so that the minimal number of mutations required to reach a $D_2$ structure from zero initial conditions is the same as that required to reach a $C_4$ structure, the redundancy available to $D_2$ genomes (which may contain, for example, one unpaired heterointeraction in addition to their homointeractions) gives $D_4$ a higher search space volume than that of the less redundant $C_4$ structures.  We can also map out all the pathways between different phenotypic states, as shown in Fig.~\ref{s18}.   

To study the dynamic effects of the structure of search space, we simulated a population of $10^5$ random walkers in genome space. Each walker started from zero initial conditions, and then took mutational steps until a genome encoding one of the two tetrameric states was reached. A mutational step involved an application of the mutation operator from a GA, rather than enforcing exactly one mutation per step. Walks were terminated and ignored if they reached the UND state (something that mirrors what might happen in nature where this usually would be lethal for the organism).

  A similar random walker analysis is possible in phenotype space, on the network in Fig. \ref{s18} a) ii). Here, each random walker occupies a node in the network, and may, at each time-step, undergo a transition between nodes according to the weight of the connecting edge. A population of walkers was initialised at the monomer node and allowed to walk, with UND encounters being terminated and ignored. The results of both these walker simulations are shown in Table \ref{nsiff}.

\begin{table}
\begin{tabular}{  c  c  c  c  c  c }
\hline\hline		
  & UND & Monomer & Dimer & $D_2$ & $C_4$ \\
\hline
$n_{si} = 2$ & $1\,214$ & $994$ & $1\,488$ & $272$ & $128$ \\
$n_{si} = 3$ & $1\,829$ & $431$ & $1\,212$ & $552$  & $72$ \\
$n_{si} = 4$ & $2\,510$ & $146$ & $736$ & $672$ & $32$ \\
\hline\hline
\end{tabular}
\caption{Number of genomes in the $\mathcal{S}'_{1,8}$ search spaces that encode different structures.}
\label{ngen}

\end{table}

To test the effect of a dimer intermediate on the probability of obtaining a $D_2$ or a $C_4$ structure, we also used a GA to run an evolutionary simulation.
We employed two different fitness functions, representing two situations: dimers possessing either no fitness advantage or a large fitness advantage over monomers. As the only possible phenotypes in this landscape are UND and $s = 1$, $s = 2$, $s = 4$, we represent a fitness function with the values awarded to these four cases respectively. Fitness function A gives no advantage to dimer formation: $F(\mbox{UND}) = 0, F(1) = 0.1, F(2) = 0.1, F(4) = 1$. Fitness function B gives a large fitness advantage to dimer formation: $F(\mbox{UND}) = 0, F(1) = 0.1, F(2) = 0.9, F(4) = 1$. We ran $10^4$ simple GAs for each case, with $N=80$ and $\mu L = 0.5$, and measured the proportion of times a run discovered (rather than adapted to) either a $C_4$ or a $D_2$ phenotype.  Table \ref{nsiff} shows the results of simulations  with these fitness functions. 

We then introduced an evolutionary bias towards homointeractions by allowing more colours to self-interact. To this end, we first include $1 \leftrightarrow 1$ (giving $n_{si} = 3$) and then $5 \leftrightarrow 5$ bonds (giving $n_{si} = 4$). This addition of homointeractions changes the evolutionary landscape dramatically (see Fig. \ref{s18} b)). The distribution of phenotypes is shown in Table \ref{ngen}. The number of UND genotypes is observed to increase with $n_{si}$, due to the greater number of genomes that encode extended, unbound structures in systems with large numbers of self-interactions.

\begin{table}
\begin{tabular}{ c c c c c c }
\hline\hline		
  & SSP & GW & PW & GA, FF A & GA, FF B \\
\hline
$n_{si} = 2$ & 0.68 & 0.58 & 0.48 & 0.55 & 0.67 \\
$n_{si} = 3$ & 0.88 & 0.75 & 0.67 & 0.69 & 0.80 \\
$n_{si} = 4$ & 0.95 & 0.86 & 0.75 & 0.96 & 0.98 \\
\hline\hline  
\end{tabular}
\caption{Proportion of random walker and GA simulations on $\mathcal{S}'_{1,8}$ that result in a $D_2$ structure being discovered before a $C_4$ structure. Columns are Search Space Proportion (SSP), defined as the number of genomes encoding $D_2$ structures divided by the total number of tetramer genomes, Genotype Walker (GW), Phenotype Walker (PW), and GAs, with FF denoting fitness function, as described in the text. $n_{si}$ is the number of self-interacting colours in the rule set. GAs were run with $N = 80, \mu L = 0.5$.}
\label{nsiff}

\end{table}

Table \ref{nsiff} shows a comparison of the $D_2:C_4$ ratio expected from search space structure with results for walker and GA simulations on these systems.  A number of trends can be observed in Table \ref{nsiff}. 
Although the interactions are chosen so that the minimum number of mutations required to reach a $D_2$ structure from zero initial conditions is the same as that for a $C_4$ structure, $D_2$ structures appear more frequently, which is commensurate with the fact that they occupy a larger proportion of search space.  

However,  the proportion of runs that first discover $D_2$ structures is lower than expected from the search space structure for genotype walkers, and lower still for phenotype walkers.  The slightly lower proportion for genotype walkers is due to the starting point of the simulations: monomers, of all possible phenotypes, display the highest transition probability to $C_4$ structures, so $C_4$ discovery is more likely from the zero initial conditions we employ (encoding a monomer) than from a random start point. 

The significantly lower $D_2$ proportion from phenotype walkers is due to the shorter length of the monomer $\rightarrow$ tetramer pathways, which requires only one transition, whereas monomer $\rightarrow$ dimer $\rightarrow$ $D_2$ requires two. In this case, the phenotype representation has masked the genetic detail whereby the minimal number of steps required to reach $D_2$ and $C_4$ structures from zero initial conditions are identical. The steps in the minimal monomer $\rightarrow$ $C_4$ pathway involve one neutral monomer step (0000 $\rightarrow$ 0002) and one phenotype-changing monomer $\rightarrow C_4$ step (0002 $\rightarrow$ 0062), whereas both steps to reach a $D_2$ structure are phenotype-changing (0000 $\rightarrow$ 0003 $\rightarrow$ 0043). The observed difference is an illustration of the influence of a complex genotype-phenotype map. In this case, information is lost when mutational steps across a neutral network are disregarded.

The proportion of GA runs with fitness function A that identify a $D_2$ structure is lower than expected in comparison to the genotype walkers for $n_{si} = 2$ and $n_{si} = 3$. This difference arises from the different amounts of time required for a GA to identify $D_2$ and $C_4$ structures. For $n_{si} = 2$ and $3$, it is observed that the mean discovery time for $C_4$ structures is lower than the mean discovery time for $D_2$ structures. A GA reports the structure it first discovers, whereas a set of genotype walkers reports the proportion of structures discovered regardless of the relative time taken to reach these structures. The lower $C_4$ mean discovery time for low $n_{si}$ GAs therefore results in more $C_4$ structures being reported than in the genotype walkers. For $n_{si} = 4$, the mean discovery time for $D_2$ structures is lower than that for $C_4$ in GAs, reflected in the higher observation of $D_2$ structures in these GA simulations. Note that if the GWs were run in parallel sets, and the set was stopped at the first discovery of a tetramer, this would also favour $C_4$ for $n_{si} = 2, 3$ and $D_2$ for $n_{si} = 4$. 

Another effect that acts to change the expected $D_2:C_4$ ratio arises from UND structures. In a GA, genomes encoding UND structures will be replaced (due to their zero fitness) by a copy of another genome chosen by selection. This replacement genome will be either a monomer or a dimer, according to the current state of the GA population. As $n_{si}$ increases, or if fitness function B is used, the population becomes more likely to contain dimers, due respectively to their increased presence in search space and their increased fitness. If UND genotypes are replaced by monomers, $C_4$ discovery will be more likely (the case at low $n_{si}$). If they are replaced by dimers, $D_2$ discovery will be more likely (the case at high $n_{si}$).


Another noticeable result is that conferring a fitness advantage to dimers increases the proportion of $D_2$ structures discovered in GAs. This increase is due to selection favouring dimers in the evolving population, from which situation the dimer $\rightarrow D_2$ transition is most likely.



The above GA results concern the discovery of tetramers rather than adaptation of the population to tetramers. When adaptation was considered, the $n_{si} = 2$ and $3$ trends remained very similar. The $n_{si} = 4$ system became 100\% dominated by $D_2$ tetramers, as the genomes encoding $C_4$ structures in this system were individual and isolated. In other words, they exhibit low phenotypic robustness and adaptation proved impossible with such small neutral network sizes.  
 
We note that the coarse-grained nature of our model greatly simplifies the description of protein surfaces. In proteins, interacting sites consist of multiple amino acid residues, rather than a single colour type as we employ. Point mutations in reality will normally alter not more than one constituent amino acid of a bonding site, rather than entirely changing the bonding characteristics of an interaction site. In addition, the spatial structure of protein complexes is vastly more complicated than the simple 2D tile geometry we employ here. However, this simple system nonetheless displays interesting dynamic behaviour.  We show that favouring homo-interactions in the search space, and favouring dimers in the fitness function, can both significantly enhance the proportion of $D_2$ tetramers over $C_4$ tetramers.  By performing a complete enumeration of the the fitness landscape, we can also uncover some subtle questions related to the underlying structure of the landscape.   For example, considering only the phenotype structure can mask important genotypic structure that influences the evolutionary dynamics.

\section{\label{secconc}Conclusions}

We have studied the evolutionary dynamics of self-assembling  polyominoes.      We focussed on deterministic self-assembly -- where a given rule set always leads to the same polyomino structure -- because an analogy can be made with  monodisperse self-assembly seen in nature, for example in protein quaternary structure.     

Although our model is simple enough to be easily tractable with modest computational resources, it exhibits rich evolutionary behaviour that is linked to its non-trivial genotype-phenotype mapping.    The evolutionary dynamics can be viewed as a search performed by a population of individuals on a complex fitness landscape. An advantage of the polyomino system is that in some cases this landscape can be fully enumerated and classified in terms of adjacent structures and the transitions between them.   Such information helps explain some of the detailed behaviour observed in GA simulations.  Properties like robustness and evolvability \cite{wagner2005robustness}  can easily be extracted from the fully enumerated landscapes.  

We also investigated the effect of changing the mutation rate, the population size, and the size of the search space on adaptation and discovery times for the evolution of certain classes of polyominoes.  We find that there is an optimal, intermediate mutation rate value for adaptation.    For smaller $\mu L$ the system takes longer to discover the desired phenotypes, whereas for larger $\mu L$ the mutational entropy prevents it from adapting to the right phenotype.

For smaller spaces and larger populations the discovery time keeps decreasing with increasing mutation rate, but for larger spaces, there is also an optimal mutation rate for the discovery time. This effect can be cast into the language of \emph{exploration and exploitation}~\cite{march1991exploration,eiben1998evolutionary, holland1992adaptation}. For low $\mu L$, the system can only take small steps across the search space, leading to confinement of the gene pool around fitness optima \cite{wrig32}, low diversity, and slow {\em exploration} of surrounding space. At high $\mu L$, the system de-correlates very quickly, reducing its ability to  {\em exploit} beneficial mutants through further small changes, raising diversity to almost the level expected for a randomised population. The search's hill-climbing ability is decreased as large steps randomise the gene pool very quickly.   

The modelling of evolutionary processes with genetic algorithms is complicated by the fact that the number of parameters that can be varied is very large.  One advantage of our polyomino system is that the effects of varying the GA parameters can be easily quantified.   We studied some popular parameter choices, and argue that, for example, the use of random initial conditions or elitism may not be the most biologically relevant way to parameterise a genetic algorithm.


Finally, we studied the evolution of polyomino tetramers, inspired by recent work on the structure and evolution of homomeric protein tetramers~\cite{levy2008assembly,villar2009self}.  In nature there is a strong preference of $D_2$ over $C_4$ symmetries, and we show that both an increase in the probability of homointeractions as well as a fitness advantage of dimeric intermediates can strongly favour the formation of $D_2$ symmetry.   Our simplified model shows that the outcome of evolutionary dynamics is affected by the topology of the search space, including emergent properties like phenotypic robustness.

\bibliography{polyevo4}{}

\end{document}